\documentclass[aps,prl,groupedaddress,superscriptaddress,showpacs,reprint,nofootinbib]{revtex4-2}
\usepackage{amsmath}
\usepackage{amsfonts}
\usepackage{bbold}
\usepackage{bm}

\usepackage[dvipsnames,svgnames,table]{xcolor}
\definecolor{OliveGreen}{cmyk}{0.64,0,0.95,0.40}
\usepackage[colorlinks=true,
            linkcolor=blue,
            urlcolor=blue,
            citecolor=OliveGreen,          
            bookmarks=true,
            bookmarksnumbered=true,
            breaklinks=true,
            pdfpagemode=FullScreen,
            pdfstartview=FitBH]{hyperref}
\usepackage{graphicx}
\usepackage{stackengine}
\usepackage{color}
\usepackage{subfigure}

\usepackage{helvet}
\usepackage{times}
\usepackage{feynmp-auto}

\usepackage[inline]{enumitem}
\usepackage{orcidlink}
\usepackage{lipsum}
\usepackage{comment}
\usepackage[normalem]{ulem}
\usepackage{multirow}

\usepackage{makecell}

\definecolor{lime}{HTML}{A6CE39}
\DeclareRobustCommand{\orcidicon}{%
	\begin{tikzpicture}
	\draw[lime, fill=lime] (0,0) 
	circle [radius=0.16] 
	node[white] {{\fontfamily{qag}\selectfont \tiny ID}};
	\draw[white, fill=white] (-0.0625,0.095) 
	circle [radius=0.007];
	\end{tikzpicture}
	\hspace{-2mm}
}

\foreach \x in {A, ..., Z}{\expandafter\xdef\csname orcid\x\endcsname{\noexpand\href{https://orcid.org/\csname orcidauthor\x\endcsname}
	{\noexpand\orcidicon}}
}

\definecolor{OliveGreen}{cmyk}{0.64,0,0.95,0.40}

 \newcommand{\fer}[1]{{\color{black}  #1}}

 \newcommand{\op}[1]{{\color{black}  #1}}
 \newcommand{\ac}[1]{{\color{black} #1}}

 \newcommand{\risk}[1]{}

\begin{document}

\title{Alleviating the present tension between T2K and NO$\nu$A with nonstandard neutrino interactions}

\author{Adriano Cherchiglia \orcidD{}}
    \affiliation{Instituto de F\'isica Gleb Wataghin - Universidade Estadual de Campinas (UNICAMP), {13083-859}, Campinas SP, Brazil}
    \affiliation{Departamento de F\'isica Te\'orica y del Cosmos, Universidad de Granada, Campus de Fuentenueva, E–18071 Granada, Spain}

\author{Pedro Pasquini \orcidF{}}
    \affiliation{Department of Physics, University of Tokyo, Bunkyo-ku, Tokyo 113-0033, Japan}
\affiliation{Instituto de F\'isica Gleb Wataghin - Universidade Estadual de Campinas (UNICAMP), {13083-859}, Campinas SP, Brazil}
\author{O. L. G. Peres \orcidC{}}
    \affiliation{Instituto de F\'isica Gleb Wataghin - Universidade Estadual de Campinas (UNICAMP), {13083-859}, Campinas SP, Brazil}

\author{F. F. Rodrigues \orcidE{}}
    \affiliation{Instituto de F\'isica Gleb Wataghin - Universidade Estadual de Campinas (UNICAMP), {13083-859}, Campinas SP, Brazil}
    \affiliation{Institute of High Energy Physics, 19B Yuqhan Road, Beijing, 100049,
    China}
 
\author{R. R. Rossi \orcidA{}}
 	\affiliation{Instituto de F\'isica Gleb Wataghin - Universidade Estadual de Campinas (UNICAMP), {13083-859}, Campinas SP, Brazil}
  	\affiliation{Instituto de Física Corpuscular, Universitat de València, E-46980, Valencia, Spain}	

\author{E. S. Souza \orcidB{}}
    \affiliation{Instituto de F\'isica Gleb Wataghin - Universidade Estadual de Campinas (UNICAMP), {13083-859}, Campinas SP, Brazil}
	
	\date{\today}
	
\begin{abstract}
Since neutrino oscillation was observed, several experiments have been built to measure its parameters. 
NO$\nu$A and T2K are two long-baseline experiments dedicated to measuring mainly the mixing angle $\theta_{23}$, the charge-parity conjugation phase $\delta_{\rm CP}$, and the mass ordering. However, 
there is a tension in current data. The T2K allowed region is in conflict with the region allowed by NO$\nu$A.
We propose a nonstandard charged current interaction (\ac{CC-}NSI) in neutrino production to relieve this tension. 
The \ac{CC-}NSI is computed through quantum field theory (QFT) formalism, 
where we derive perturbative analytical formulae considering \ac{CC-}NSI in the pion decay.
Within this new approach, we can alleviate NO$\nu$A and T2K tension for a \ac{CC-}NSI complex parameters of order $10^{-3}$. We show the new phase has a degeneracy to the Dirac CP phase of the form $\delta_{\rm CP} \pm \phi= 1.5\pi$ being a possible source of violation of charge-parity symmetry.
\end{abstract}
\maketitle

\textit{Introduction}.--- The neutrino oscillation phenomenon provides evidence of physics beyond the Standard 
Model. Since its discovery, several experiments have measured neutrino oscillation 
parameters \cite{Esteban:2020cvm, *Esteban:2024eli, deSalas:2020pgw,Capozzi:2021fjo}. 
One not yet measured is the charge-parity (CP) conjugation phase 
$\delta_{\rm CP}$ that quantifies the asymmetry between particle and anti-particle. The two long-baseline accelerator 
experiments, NO$\nu$A and T2K, were designed to measure this parameter.

\ac{The NO$\nu$A and T2K allowed parameter regions are in tension for sometime~\cite{Nizam:2018got,Kelly:2020fkv,Esteban:2020cvm,Kelly:2020fkv,Capozzi:2021fjo}, and it persists in new data~\cite{NOvA:2021nfi,T2K:2023smv}}.
In the standard three-neutrino oscillation scenario, each individual experiment has a preference for normal ordering, while their combination indicates a preference for inverted ordering. These results could indicate physics beyond the Standard Model \ac{(BSM)}. Numerous studies have been dedicated to explaining this tension, exploring various new 
physics scenarios \op{such as non-unitary mixing matrix~\cite{Dutta:2016vcc,Miranda:2019ynh,Yu:2024nkc}, 
neutral current NSI in propagation~\cite{Capozzi:2019iqn,Chatterjee:2020kkm,*Chatterjee:2024kbn,Denton:2020uda,Majhi:2022wyp}, 
light and very light sterile neutrinos~\cite{Chatterjee:2020yak,deGouvea:2022kma}, 
Lorentz violation~\cite{Rahaman:2021leu,*Rahaman:2022rfp}
and dark photon scenarios~\cite{Lin:2023xyk,*Alonso-Alvarez:2024wnh,Konwar:2024nwc}.}

We propose a novel approach that includes non-standard interactions in neutrino
production specifically via pion decay. 
By adopting an effective field theory approach \cite{Falkowski:2019xoe,Falkowski:2019kfn, Du:2020dwr,Falkowski:2021bkq,Chaves:2021kxe,Du:2021rdg,Breso-Pla:2023tnz,Kopp:2024yvh,Cherchiglia:2023aqp,Coloma:2024ict,Breso-Pla:2025pds,Kling:2025zsb}, we can straightforwardly modify the rate of pion decay to include
these non-standard interactions during production. 
We have derived for the first time a perturbative analytical expression for a neutrino oscillation in matter
considering this new interaction at the source. \ac{As there are stringent bounds coming from pion decay experiments~\cite{ParticleDataGroup:2024cfk}, one may wonder how much room there is for BSM physics to still alleviate the tension. We will show that our proposal not only decreases tension, but also provides a better fit to the present data.}

The new coupling constant may be complex, 
which introduces a new charge-parity violation phase. 
We investigated the interaction between the two phases: one originating from the Pontecorvo-Maki-Nakagawa-Sakata (PMNS) neutrino mixing matrix~\cite{Maki:1962mu,Pontecorvo:1957cp} and the other from the effects of
the new interaction.

In this Letter, we demonstrate that the tension is alleviated even if only one new complex parameter can be non-zero. We have determined that the absolute value of the new interaction parameter is $(10^{-4}-10^{-3})\times G_{\rm F}$, the Fermi constant.

\textit{New physics in neutrino sector from an EFT perspective}.--- We consider the non-standard interactions
on neutrino production essentially following the formalism introduced in \cite{Falkowski:2019xoe,Falkowski:2019kfn, Falkowski:2021bkq,Kopp:2024yvh}. 
The new physics is described by Wilson coefficients of four-fermion effective interactions between 
neutrinos ($\nu_{\beta}$), 
charged leptons ($l_{\alpha}$)
and quarks ($q_{i}$), $\sim \overline q_i \Gamma_A^{ij} 
q_j\bar{\ell}_{\alpha}\Gamma_A'^{\alpha\beta} P_L 
\nu_{\beta}$, where $i,j=u,d,c,s, \dots$ and $\alpha,\beta=e,\mu,\tau$.
The index $A$ corresponds to the Lorentz indices of the interaction. All possible combinations of the vertex structure are encoded in $\Gamma, \Gamma'$.
Typically, neutrinos are produced via pion decay, and only
vector, axial, and pseudo-scalar couplings with $q_{i} = u$, $q_j = d$ contribute. In this Letter we only consider the latter, which could be generated by models containing extra pseudo-scalars, for instance. We emphasize that, by suitable normalization, our results can be translated to the other cases as well. Explicitly, the effective Lagrangian will then contain the term~\cite{Falkowski:2019xoe,Falkowski:2019kfn, Falkowski:2021bkq,Kopp:2024yvh}
\begin{align}
    \mathcal{L}_{\rm P}
\supset
    \sqrt{2}\, G_{\rm F} V_{\rm ud}^{\rm CKM}
    \epsilon_{\alpha \beta}
    \left(\bar{u} \gamma^5 d \right)\left(\bar{\ell}_{\alpha}P_L \nu_{\beta} \right)
+\mathrm{h.c.} ~ ,
\label{eq:lagrangian}
\end{align}
where $V^{\rm CKM}$ is the Cabibbo-Kobayashi-Maskawa (CKM) matrix~\cite{Cabibbo:1963yz, *Kobayashi:1973fv}
$G_{\rm F}$ is the Fermi constant. Furthermore, $\epsilon_{\alpha \beta}$ are complex Wilson coefficients that describe the magnitude of the new interaction relative to the weak interaction. The new interaction in Eq.~(\ref{eq:lagrangian}) creates another vertex for neutrino production beyond the traditional one~\cite{sup}.
With the new vertex, the total matrix element is a combination of the standard model amplitude ($\mathcal{A}_L^{\rm S}$) and the new physics amplitude ($\mathcal{A}_{P}^{\rm S}$),
\begin{align}
  \mathcal{M}_{\alpha k}^S 
= 
  U_{\alpha k}^*
  \mathcal{A}_L^S 
+
  \left[
        \epsilon \, U
  \right]_{\alpha k}^* 
  \mathcal{A}_{P}^S ~ .
\label{eq:amplitudes}
\end{align}
The upper index, $S$, for the source indicates that the process occurs only in production, since the detection process induced by a pseudo-scalar is extremely suppressed
\cite{Falkowski:2019xoe,Falkowski:2019kfn,Falkowski:2021bkq}. Therefore, there are no relevant effects on the detection of the new interactions.
It should be emphasized that neutrino mass eigenstates are encoded exclusively in PMNS mixing matrices \cite{Maki:1962mu,Pontecorvo:1957cp}, so that amplitudes $A_{L/P}^{\rm S}$
depend solely on the neutrino flavor. Notice that the
off-diagonal terms of $\epsilon_{\alpha\beta}$ violate the lepton flavor number.

\begin{figure}[!htp]
\includegraphics[scale=0.15]{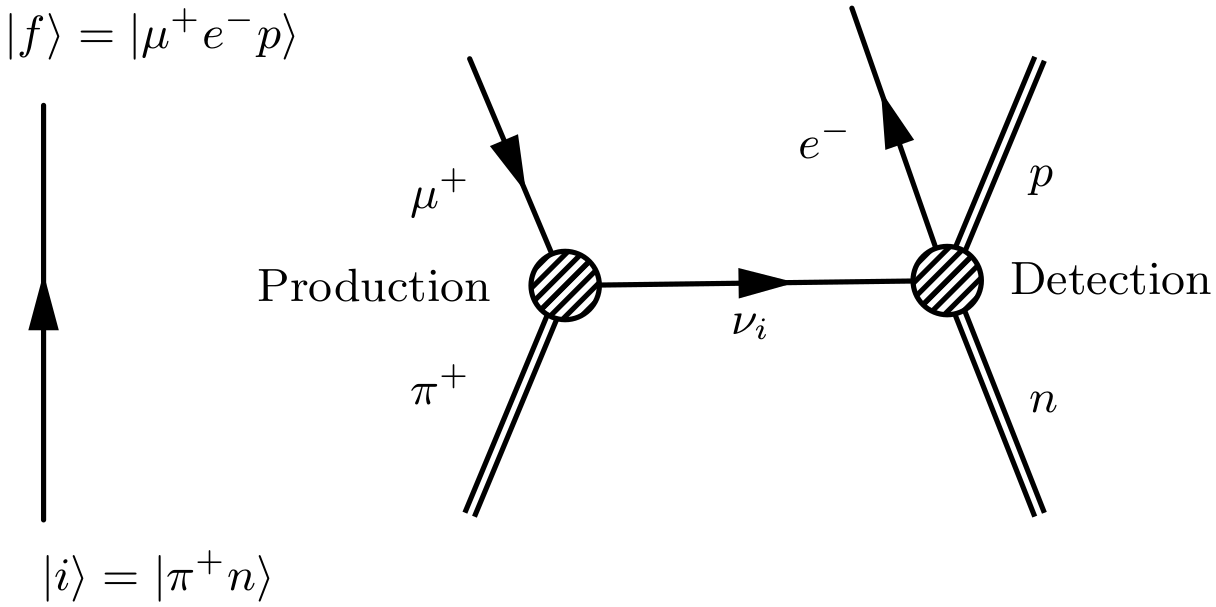}
    \caption{Quantum field theory computation of neutrino oscillation probability~\cite{Falkowski:2019xoe,Falkowski:2019kfn, Du:2020dwr,Falkowski:2021bkq,Chaves:2021kxe,Du:2021rdg,Breso-Pla:2023tnz,Kopp:2024yvh,Cherchiglia:2023aqp,Coloma:2024ict,Breso-Pla:2025pds,Kling:2025zsb}.}
\label{eft-probability}
\end{figure}

\textit{Neutrino event rate in the QFT formalism}.--- The event rate is the physical observable in neutrino
oscillation experiments. In the formalism of Quantum Field Theory (QFT),  neutrino production, 
propagation, and detection are considered \ac{a} single process.
Therefore, the neutrino oscillation is quantified by a single tree diagram, as illustrated in Figure~(\ref{eft-probability}) by the decay $\pi^{+} \rightarrow \mu^{+} + \nu$ (production) followed by detection $\nu~+~n~ \rightarrow~ p~ +~ e^{-}$. The time direction is from bottom to top. In the production and detection processes, the initial states are the pion and neutron. The detected particles (e.g., charged leptons and protons) are regarded as final states~\cite{sup}.
The neutrino participates in the process as an intermediate state, where the uncertainties of the initial state result
in the superposition of massive neutrino states
~\cite{Grimus:1996av}.

In this formalism, the neutrino event rate, including \ac{CC-}NSI in production is~\cite{Falkowski:2019xoe,Falkowski:2019kfn,Falkowski:2021bkq}
\begin{align}
    R_{\alpha \beta}^{\rm CC-NSI}
= &   
    \kappa
    \textstyle{\sum_{k j}}
        e^{ -i \Delta_{kj}
 }
    \mathcal{U}_{\beta k} \mathcal{U}_{\beta j}^{*}
    \nonumber\\
&    
   \times 
   \int 
        d\Pi_{S} 
        \mathcal{M}_{\alpha k}^S 
        \overline{\mathcal{M}}_{\alpha j}^S
   \times 
   \int 
        d\Pi_{D} 
        |\mathcal{A}_{L}^D|^{2} ~ ,
\label{eq:event_rate}    
\end{align}
where $\alpha$ and $\beta$ denote produced and detected flavor states, respectively, $\kappa$ is a constant that includes the kinematical factors and target size, 
$\Delta_{kj}\equiv  \frac{\Delta m_{kj}^2 L}{2E_\nu}$, with $E_\nu$ being the neutrino energy, 
$L$ the source-detector distance, 
and  
$\Delta m_{kj}^2 \equiv m_k^2 - m_j^2$ the neutrino mass squared difference and the amplitude $\mathcal{M}_{\alpha k}^S $  
is given in Eq.~(\ref{eq:amplitudes}). The integrals are over the phase space elements for source ($S$) and detection ($D$).
We denote by $\mathcal{U}$ the PMNS mixing matrix~\cite{Maki:1962mu,Pontecorvo:1957cp} in constant matter~\cite{Mikheev:1986wj,Wolfenstein:1977ue}.

The events rate Eq.(\ref{eq:event_rate}) is associated 
to the oscillation probability by the definition: $P_{\alpha \beta}^{\rm NSI} 
\equiv  R_{\alpha \beta}^{\rm NSI}/\phi_\alpha^{{\rm SM}} \sigma_\beta^{{\rm SM}}$, corresponding to the transition 
$\nu_{\alpha}
\rightarrow
\nu_{\beta}$.
It is conveniently written~\cite{newformula} as
{\small
\begin{align}
    P_{\alpha \beta}^{\rm CC-NSI}
\!=\!
    \sum_{k j} 
    e^{ -i 
    \Delta_{kj}}
    \!
    \big[\!
        \left(
            \mathbb{1}
            -            p_{\alpha} \epsilon
        \right)\! \mathcal{U}
    \big]_{\alpha k}^{*}
   \big[\!
        \left(
            \mathbb{1}
            - 
            p_{\alpha} \epsilon
        \right)\! \mathcal{U}
    \big]_{\alpha j} 
    \mathcal{U}_{\beta k}\mathcal{U}_{\beta j}^{*} ~ , 
    \label{eq:probability}
\end{align}
}
\noindent
where $p_{\alpha}= m_\pi/(m_\alpha(m_u+m_d))$, for example $p_{\mu} \sim 27$ and $p_{e} \sim 5500$ represents a chiral enhancement compared to the standard model rate~\cite{Guzzo:2023ayo}. \op{ This is the reason that we choose the pseudo-scalar scenario, since the standard model rate is chirality supressed.}

In the end, the effect of \ac{CC-}NSI consists of substituting the matrix $\mathcal{U}_{\alpha i}$ by $[(\mathbb{1} - p_{\alpha}\epsilon)\mathcal{U}]_{\alpha i}$.
Although we have named Eq.~(\ref{eq:probability}) as the probability because of its similarity to the traditional form,
the presence of \ac{CC-}NSI makes the expression effectively unitarity-violating. 

In order to analyze the impact of individual \ac{CC-}NSI parameters on the oscillation probability, we consider two scenarios corresponding to a new source for muon or electron neutrinos. 
In the EFT formalism, they are implemented by allowing for only one non-zero Wilson coefficient at a time, $\epsilon_{\mu e}$ or $\epsilon_{e \mu}$, respectively. For the experimental analyses of interest, the parameter $\epsilon_{\mu e}$ will modify the signal and $\epsilon_{e \mu}$ will affect the background.
In the following, we will discuss the $\epsilon_{\mu e}$ scenario to exemplify the perturbative formalism. Because the initial state in pion decay is muonic neutrino, we need to calculate the probability $P_{\mu \beta}$.
We write for the first time an analytical formula
for Eq.~(\ref{eq:probability}) in terms of the
evolution operator $S^{\rm OSC} \equiv  e^{-i H t}$ for the neutrino Hamiltonian, defined in a standard oscillation scenario~\cite{newformula}.
Therefore,
\begin{align}
P_{\mu \beta}^{\rm CC-NSI}
    =
    \left|
        S_{\beta \mu}^{\rm OSC} 
        - 
        p_\mu 
        \epsilon_{\mu e}^{*}
        S_{\beta e}^{\rm OSC}
    \right|^2,
\label{eq:prob_Smatrix}
\end{align}
where the complex coefficient is explicitly
$\epsilon_{\mu e} \equiv
|\epsilon_{\mu e}|
e^{i\phi_{\mu e}}.
$
The advantage of writing the probability above 
is that there is in the literature  
the analytical expression for $S^{\rm OSC}$ 
with matter effects~\cite{Asano:2011nj}. It can also be straightforwardly generalized to other \ac{CC-}NSI scenarios, and other conversion/survival rates.

The most important equation of this paper is the $\nu_\mu\to \nu_e$ probability with \ac{CC-}NSI, using an analytical expression {\it in matter}. We have derived it employing a perturbative approach~\cite{Asano:2011nj}, where the leading terms are given by
\begin{widetext}
\begin{align}
\nonumber
    P_{\mu e}^{\rm NSI} 
&=
    4 \frac{ s^2_{13} s_{23}^{2} }{ (1 - r_{a})^2 }
    \sin^2 \frac{ (1 - r_{a}) \Delta L }{ 2 } 
+ 
     \frac{
    8 
    J_{r} r_{\Delta} }{ r_{a} (1 - r_{a}) }
    \cos 
    \left( 
        \delta_{\rm CP} 
    +\frac{ \Delta L }{ 2 }
    \right)
    \sin \frac{ r_{a} \Delta L }{ 2 }
    \sin \frac{ (1 - r_{a}) \Delta L }{ 2 }
\\&
+   p_\mu^2 |\epsilon_{\mu e}|^2+
    4 p_\mu
    |\epsilon_{\mu e}|
    \frac{s_{13}s_{23}}{1-r_{a}}
     \sin{\left(\frac{(1-r_{a})\Delta L}{2}\right)}
    \sin{\left(\delta_{\rm CP} 
    - \phi_{\mu e}
    + \frac{(1-r_{a})\Delta L}{2}\right)} 
    + \op{{\cal O}(r_\Delta s_{13},s_{13}^3)}
    ~ .
\label{eq:prob_dominante}
\end{align}
\end{widetext}
From the phenomenological nature of the parameters 
$
r_{\Delta}~\equiv~
\Delta m_{21}^2/\Delta m_{31}^2 
\simeq \zeta
$
and
$
\sin \theta_{13} 
\simeq 
\sqrt{\zeta}
$
, with 
$
\zeta
\sim
\mathcal O(10^{-2})$. 
We also define $\Delta = \Delta m_{31}^{2}/2E_{\nu}$, 
$L$ is the distance between the source and detector, 
$r_{a} = a/\Delta$ 
with $a=\sqrt{2}G_{\rm F} N_e$ being the matter potential
and the \op{reduced} Jarskolg factor \cite{Jarlskog:1985ht, *Jarlskog:1985cw}
$J_{r}~=~c_{12}s_{12}c_{23}s_{23}s_{13}$ in shorthand notation
$s_{ij}=\sin{\theta_{ij}}$ and
$c_{ij}=\cos{\theta_{ij}}$.
The probability of antineutrino retains the form of Eq.~(\ref{eq:prob_dominante}) with the replacements
$
\delta_{\rm CP} \rightarrow -\, \delta_{\rm CP}
$, 
$
\phi_{\mu e} \rightarrow -\, \phi_{\mu e}
$ 
and 
$
a \rightarrow -\, a ~ .
$

The analytical formulae are very useful to identify the sources of CP violation. In the standard oscillation scenario, we recall that the survival probability ($P_{\alpha\alpha}=|S_{\alpha\alpha}^{\rm OSC}|^2$ for neutrinos of flavor $\alpha$) is a CP-even quantity. Thus, CP-odd effects can only come from processes involving the conversion of flavor between neutrinos (given by $P_{\beta\alpha}=|S_{\alpha\beta}^{\rm OSC}|^2$ with $\beta\neq\alpha$). 
In the presence of \ac{CC-}NSI, this reasoning does not hold, as can be easily checked by considering $\beta=\mu$ in Eq.~\eqref{eq:prob_Smatrix}. The case with $\beta=e$ is even more instructive. First, it follows directly from Eq.~\eqref{eq:prob_Smatrix} that the terms quadratic dependent on $|\epsilon_{\mu e}|$ will not depend on $\delta_{\rm CP}$. Secondly, the leading-order terms given in Eq.~\eqref{eq:prob_dominante} show that the presence of \ac{CC-}NSI induces a term dependent on the difference of phases
$(\delta_{\rm CP}-\phi_{\mu e})$. Since the ratio between the standard {\color{black} CP-violation} term in the first line of Eq.~\eqref{eq:prob_dominante} (\ac{proportional to $J_r$}) to the last term is of order $\zeta$, for $p_{\mu}|\epsilon_{\mu e}|~\sim~ 27~|\epsilon_{\mu e}|~>~\zeta~\sim~\mathcal~ O(10^{-2})$, the \ac{CC-}NSI term may dominate, implying that the experiment may be more sensitive to the difference $(\delta_{\rm CP}-\phi_{\mu e})$ than the standard CP phase itself. 
We will show this tendency when we present our numerical results.

Finally, the perturbative
formula is in good agreement with the exact one. 
In fact, most of the energy range of the experiments discussed here exhibits an error of less than one percent ~\cite{sup}, including the region of interest for the NO$\nu$A and T2K experiments.

\textit{Experimental and simulation details}.--- We analyze the effects of \ac{CC-}NSI in neutrino production by pion decay through two long-baseline experiments: NO$\nu$A (NuMI Off-axis $\nu_e$ Appearance) and T2K (Tokai-to-Kamioka).

The NO$\nu$A experiment~\cite{NOvA:2016kwd, *NOvA:2019cyt, NOvA:2021nfi,*NOvA:2023iam}
measures muonic neutrino disappearance and electronic neutrino appearance.
Its beam is located in the Fermilab laboratory in the United States
and it travels 810 km to the detector in Minnesota.
Neutrinos go through a matter density of 
$\rho_{{\rm NO}\nu{\rm A}} = 2.84$ g/cm$^3$.
We adopt the configuration of
$13.6 (12.5)\times 10^{20}$ protons on target (POT) for (anti-)neutrinos mode.
The mass of the target detector is 14 kt and the neutrino energy range is from 1 up to 5 GeV, with energy spectra peaked at 2.1 GeV.

The T2K experiment \cite{T2K:2011qtm, *T2K:2011ypd, *T2K:2021xwb,T2K:2023smv} also measures muonic neutrino disappearance and electronic neutrino appearance. 
The beam is produced at J-PARC laboratory in Japan and travels 295 km to the Super-Kamiokande detector. 
The density of matter in this experiment is $\rho_{\rm T2K} =  2.6$ g/cm$^3$. 
The T2K flux has {\color{black} $19.7\,(16.3) \times 10^{20}$ POT for the (anti-)neutrino mode}. The detector has a target mass of 22.5 kt, and the neutrino energy range is from 0.1 to 1.25 GeV, with energy spectra peaked at 0.6 GeV. 

{\color{black} The detector response was obtained using the GENIE software \cite{Andreopoulos:2015wxa,GENIE:2024ufm} to model the interaction cross section and the final state distributions, while the energy
reconstruction was performed through our own Monte Carlo codes.} We use GLoBES~\cite{Huber:2004ka,Huber:2007ji} {\color{black} to calculate the oscillation probabilities and obtain} the number of detected events, according to the Eq.~(\ref{eq:event_rate}), and to perform the
statistical analysis. 
We fix the solar parameters to their best-fit values~\cite{ParticleDataGroup:2024cfk} $\Delta m_{21}^2 =  7.53\times 10^{-5}{\rm eV^{2}}$ and $\sin^{2}\theta_{12}~=~0.307$, minimizing the function $\chi^2$ over all the
other relevant parameters.
We put a Gaussian prior on the reactor angle
$\sin^{2}2\theta_{13}~=~0.083\pm0.0031$
because it is well measured by other experiments~\cite{DayaBay:2022orm, *DoubleChooz:2019qbj, *RENO:2018dro}. We then present in the following 
sections, a quantitative analysis of our model, and 
the allowed region for oscillation and \ac{CC-}NSI 
parameters, for NO$\nu$A and T2K individually as well as combined.

\textit{Alleviating the T2K and NO$\nu$A tension}.--- The \ac{CC-}NSI changes the neutrino oscillation probability, as seen in Eq.~(\ref{eq:prob_dominante}). In particular, it modifies the dependence on the CP-violation parameters. In the standard oscillation scenario, a common way to illustrate the impact of the
still unknown $\delta_{\rm CP}$ parameter is to consider the idea of bievents~\cite{Minakata:2001qm,Ishitsuka:2005qi,Ribeiro:2007ud,Nunokawa:2007qh,Gago:2009ij,PhysRevD.75.033002}, in the plane of electron-neutrino versus anti-electron rates. We adopt the same idea here, but for the CC-NSI scenario. 
\begin{figure}[!htp]
\centering
\includegraphics[scale=0.27]{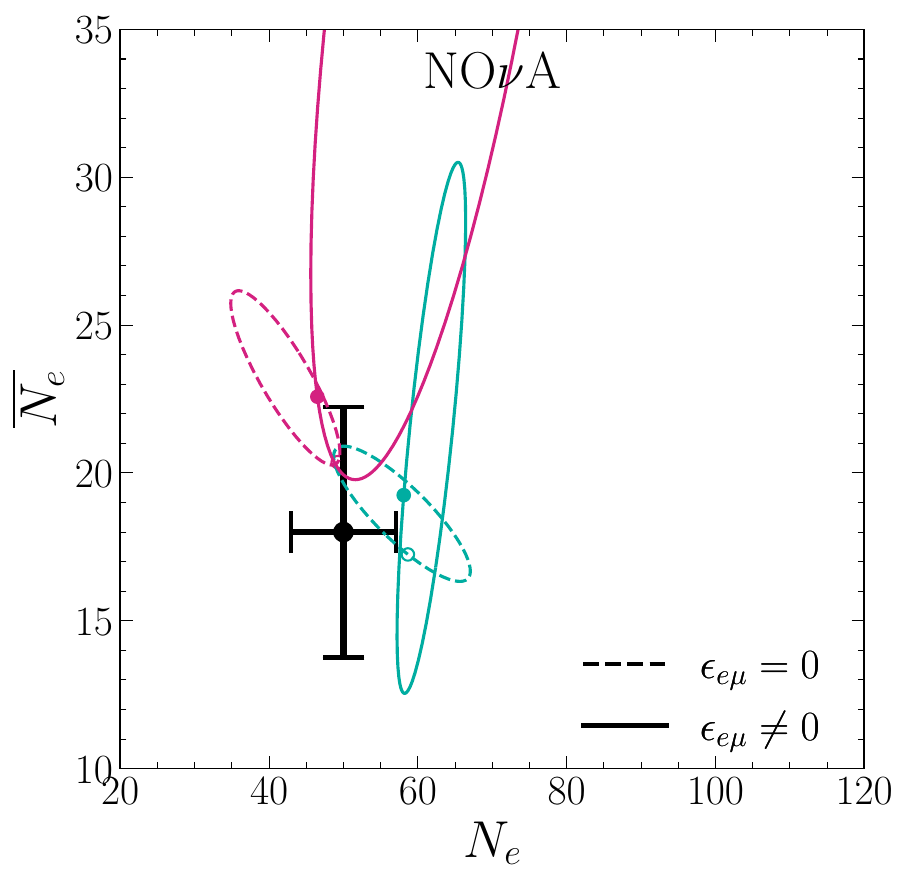}
\includegraphics[scale=0.27]{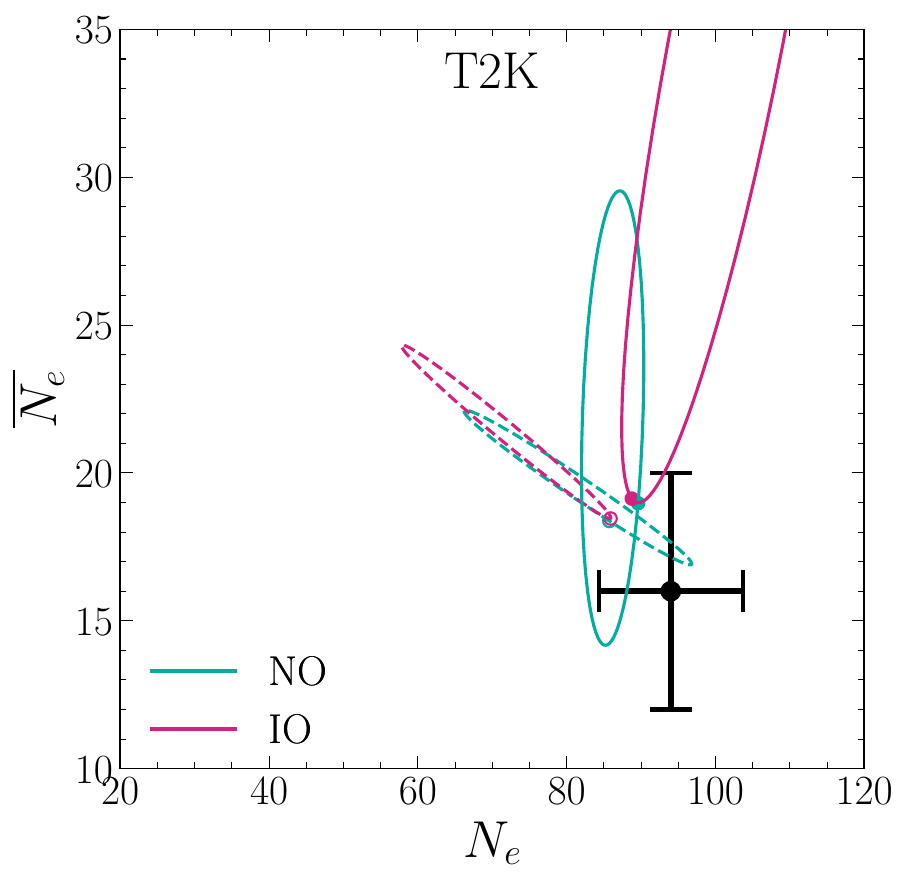}
\includegraphics[scale=0.27]{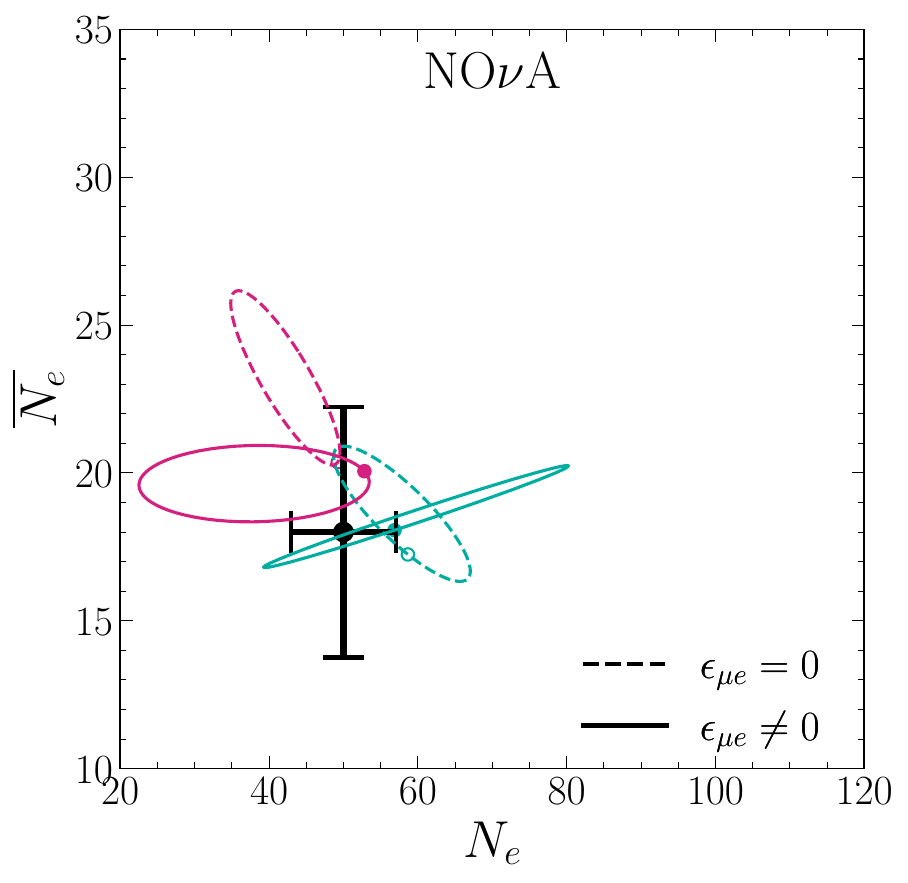}
\includegraphics[scale=0.27]{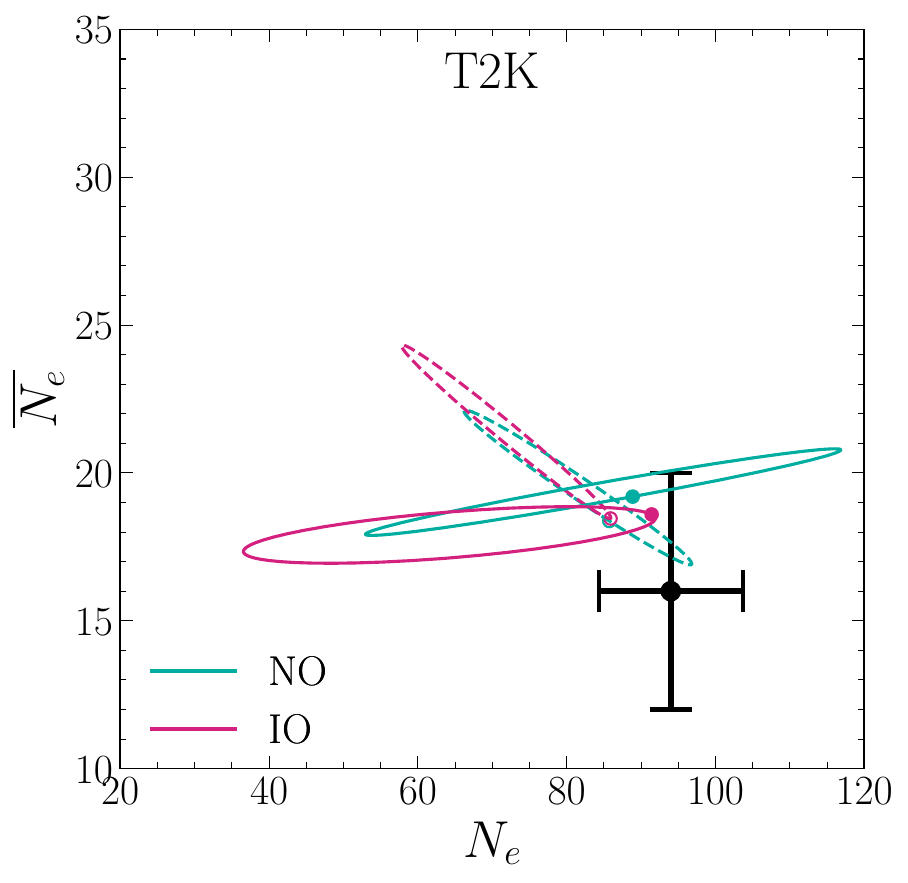}
    \caption{Bi-events plot in the plane electron number of events and anti-electron number of events for NO$\nu$A in the left panel and for T2K in the right panel, while varying $\delta_{\rm CP}$ with (solid lines) and without (dashed lines) \ac{CC-}NSI, in the NO (blue lines) and IO (pink lines). The other parameters are fixed as explained in the text. Filled (hollow) dots denote the best-fit values of solid (dashed) lines, while crosses represent estimated values of the total number of events given by each collaboration~\cite{NOvA:2021nfi,T2K:2023smv}.}
    \label{fig:elipse}
\end{figure}
\begin{table}[!htp]
\centering
\begin{tabular}{c|cc}
 \textbf{NO} & 
        $\epsilon_{e \mu }\neq 0$ & 
        $\epsilon_{ \mu e}\neq 0$  \\ \hline
$\sin^{2}\theta_{23}/10^{-1}$     
    & 5.68 (\ac{5.68})         
    & 5.69 (\ac{5.69}) \\
$\delta_{\rm CP}/\pi$             
    & 1.74 (\ac{1.35})         
    & 1.08 (\ac{1.06}) \\
$|\epsilon_{X}|/10^{-3}$       
    & 0.72 (\ac{0.70})         
    & 1.12 (\ac{1.09}) \\
$\phi_{X}/\pi$                 
    & -0.32 (\ac{0.08})       
    & -0.49 (\ac{-0.53}) \\
$(\delta_{\rm CP} \pm \phi_{X})/\pi$ 
    & 1.42 (\ac{1.43})                            
    & 1.57 (\ac{1.59}) \\
\end{tabular}
\caption{The best-fit values for the normal ordering (NO) with the \ac{CC-}NSI $\epsilon_{e\mu}$ or $\epsilon_{\mu e}$, for the combination  NO$\nu$A + T2K. The values in parentheses were obtained after including the bound from the pion decay. In the last line, the plus (minus) sign is for $\epsilon_{e\mu}$ ($\epsilon_{\mu e}$). The phases $\delta_{\rm CP}$, $\phi_{X}$, are invariant under addition of multiples of $2\pi$.}
\label{tab:BF}
\end{table}

\begin{table}[!htp]
\centering
\begin{tabular}{c||c|c||c|c||c|c}
\multirow{2}{*}{$\chi^2_{\rm min}$}
& 
 \multicolumn{2}{c||}{Standard Osc.}
&
 \multicolumn{2}{c||}{$\epsilon_{e\mu}$} 
&
  \multicolumn{2}{c}{$\epsilon_{\mu e}$}
\\ 
  & NO & IO
  & NO & IO
  & NO & IO
\\[1pt] \hline \hline
\noalign{\vskip 1.5pt}
  NO$\nu$A      &  \hspace{0.1 pt} 51.8 \hspace{0.1 pt} 
                &  \hspace{0.1 pt} 52.5 \hspace{0.1 pt} 
                &  \hspace{0.1 pt} \ac{49.3} \hspace{0.1 pt}
                &  \hspace{0.1 pt} \ac{52.1} \hspace{0.1 pt} 
                &  \hspace{0.1 pt} \ac{51.3} \hspace{0.1 pt}
                &  \hspace{0.1 pt} \ac{51.9} \hspace{0.1 pt}
\\
  T2K           &  \hspace{0.1 pt} 107.2                          \hspace{0.1 pt}
                &  \hspace{0.1 pt} 109.2 \hspace{0.1 pt}
                &  \hspace{0.1 pt} \ac{107.1} \hspace{0.1 pt}
                &  \hspace{0.1 pt} \ac{108.6} \hspace{0.1 pt}
                &  \hspace{0.1 pt} \ac{106.7} \hspace{0.1 pt}
                &  \hspace{0.1 pt} \ac{107.0} \hspace{0.1 pt}
\\
NO$\nu$A + T2K \hspace{0.1 pt}
                &  \hspace{0.1 pt} 165.9 \hspace{0.1 pt}
                &  \hspace{0.1 pt} 163.9 \hspace{0.1 pt}
                &  \hspace{0.1 pt} \ac{161.8} \hspace{0.1 pt}
                &  \hspace{0.1 pt} \ac{163.9} \hspace{0.1 pt}
                &  \hspace{0.1 pt} \ac{165.2} \hspace{0.1 pt}
                &  \hspace{0.1 pt} \ac{162.4} \hspace{0.1 pt}
\\[1.0pt] \hline \hline
\noalign{\vskip 1.5pt}
$\chi^2_{\rm \textbf{PG}}\textbf{\,/\,N}_{\rm \textbf{par}}$ 
& \hspace{0.1 pt} 7.0\,/\,4 \hspace{0.1 pt}
& \hspace{0.1 pt} 2.2\,/\,4 \hspace{0.1 pt}
& \hspace{0.1 pt} \ac{5.4}\,/\,6 \hspace{0.1 pt}
& \hspace{0.1 pt} \ac{3.2}\,/\,6 \hspace{0.1 pt}
& \hspace{0.1 pt} \ac{7.2}\,/\,6 \hspace{0.1 pt}
& \hspace{0.1 pt} \ac{3.6}\,/\,6 \hspace{0.1 pt}
\\ \hline 
\textbf{p}$_{\rm \textbf{PG} }$\textbf{-value} 
&  \hspace{0.1 pt} 14\% \hspace{0.1 pt}
&  \hspace{0.1 pt} 70\% \hspace{0.1 pt}
&  \hspace{0.1 pt} \ac{49}\% \hspace{0.1 pt}  
&  \hspace{0.1 pt} \ac{78}\% \hspace{0.1 pt}
&  \hspace{0.1 pt} \ac{30}\% \hspace{0.1 pt} 
&  \hspace{0.1 pt} \ac{73}\% \hspace{0.1 pt}
\end{tabular}
\caption{
We present the values of the $\chi^2$ minimum for the standard oscillation model and CC-NSI scenarios, for the individual datasets of NO$\nu$A and T2K and the combined analysis. The values of PG test are listed for the four free parameters in the standard oscillation scenario and six with \ac{CC-}NSI. For the non-standard scenarios, the pion decay constraints were already taken into account.}
\label{tab:analysis}
\end{table}

\begin{figure*}[!htp]
    \centering
    \includegraphics[scale=0.55]{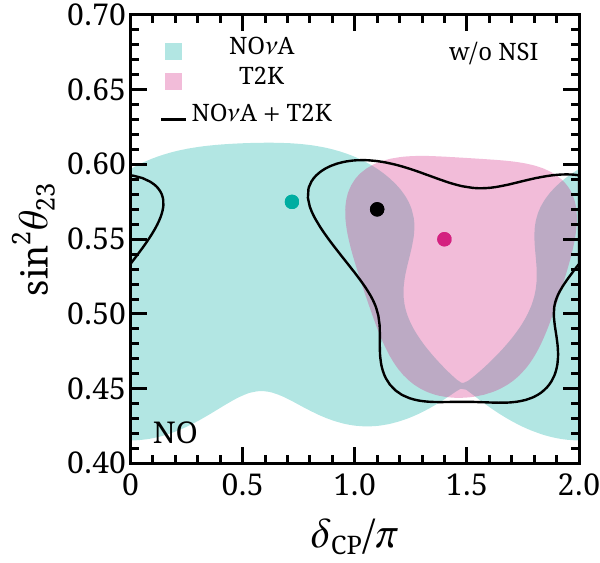}
    \includegraphics[scale=0.55]{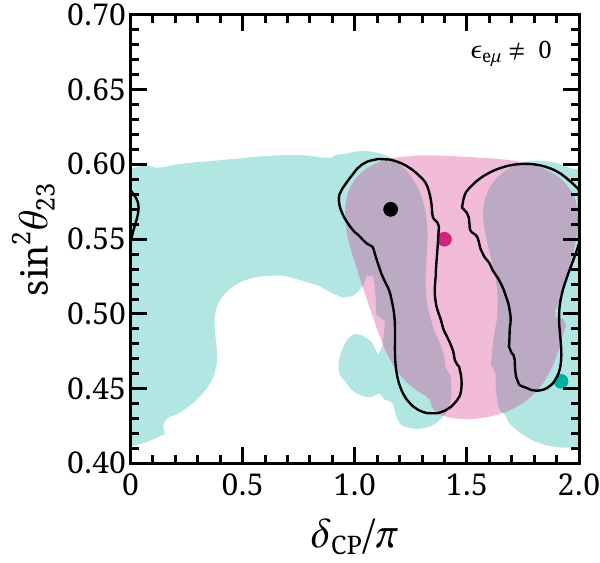}
    \includegraphics[scale=0.55
    ]{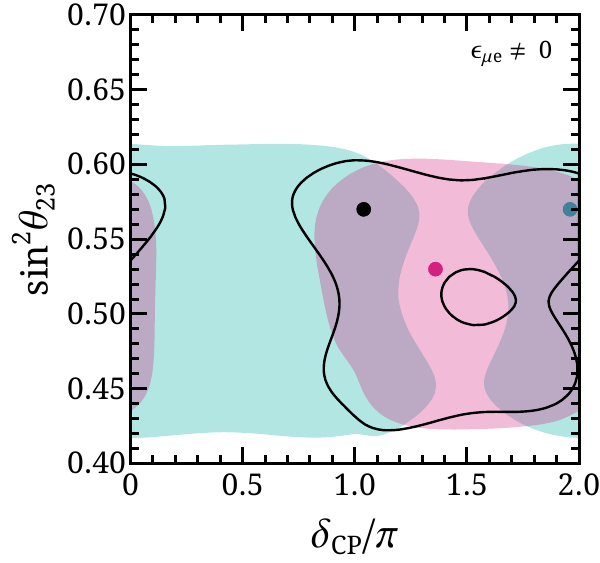}
    \caption{Allowed region for T2K (pink), NOvA (blue) and for combined analysis (black line), for NO in the $\sin^2 \theta_{23}$ vs. $\delta_{\rm CP}$ space, for 90$\%$ confidence level. In the left panel we show the standard oscillation scenario, in the middle panel  
    we show the case with $\epsilon_{e\mu} \neq 0$ and in the right panel the case with $\epsilon_{\mu e} \neq 0$.
    The dots are the respective best-fit values, see Table~(\ref{tab:BF}).
    \ac{The plots with $\epsilon_{e \mu}$ or $\epsilon_{\mu e} \neq 0$ were constructed taking into account the pion decay bounds discussed in the text.} 
   }
     \label{fig:2D_combined_V2a}
\end{figure*}

\fer{ In Figure \ref{fig:elipse} we illustrate the influence of the complex \ac{CC-}NSI, by showing the \op{total} expected number of events $N_e$ and $\bar{N}_e$
for the conversion $\nu_{\mu} \rightarrow \nu_{e}$ and $\bar{\nu}_{\mu}~\rightarrow~\overline{\nu}_{e}$, respectively}. The ellipses~\cite{Minakata:2001qm,PhysRevD.75.033002} are generated varying the value of the CP phase, 
with the remaining parameters being the combined best-fit values for NOvA and T2K.
We consider the two possible mass ordering, the so-called normal ordering (NO) and inverted ordering (IO).
In the left (right) panel we use the distance $L$  typical parameter 
for the NO$\nu$A (T2K) experiment.
We also show as dots the best fit value  for $\delta_{\rm CP}$. 
In the standard oscillation scenario, the best-fit parameters are $ \sin^{2}\theta_{23}=0.57$,
$\Delta m_{31}^2=2.50 (-2.38)\times~10^{-3}$~eV$^2$, and the CP phase
$\delta_{\rm CP}/\pi=1.10 (1.51)$, for NO (IO). In the presence of CC-NSI these 
best-fit parameters are indicated in Table~(\ref{tab:BF}) and the squared mass difference for NO is  $\Delta m_{31}^2~=~2.50 (2.47)\times 10^{-3}$~eV$^{2}$ for the scenario with CC-NSI $\epsilon_{e \mu}(\epsilon_{\mu e})$. For IO, in both scenarios the best fit-value is  $\Delta m_{31}^2~=~-2.38\times 10^{-3}$~eV$^{2}$ is 
represented by the black cross.

For the best-fit values of the 
\ac{CC-}NSI parameters, we notice the  ellipses change appreciably
even though $\epsilon_{\mu e}$ is of order $10^{-3}$.  The noticeable changes
are due to the chiral enhancement term presented in the pion decay $p_{\mu} \sim 27$, which is always multiplied by
$|\epsilon_{\mu e}|$, see Eq.~(\ref{eq:prob_Smatrix}).  In addition, the phase $\phi_{\mu e}$ introduces a new source of CP violation. The main message from Figure~\ref{fig:elipse} is: especially for NO$\nu$A, the presence of \ac{CC-}NSI allows the best-fit values (solid circles) for NO to be closer to the experimental result, in comparison to IO. As we now discuss, this will be essential to alleviate the tension between these two experiments.
Data from NO$\nu$A and T2K, the appearance of neutrinos and antineutrinos, disagree when considering the standard neutrino oscillation model. 
Each experiment individually prefers NO, but when combined, the preference is for IO.
In Figure~\ref{fig:2D_combined_V2a},
we show the allowed region with \ac{CC-}NSI in $\delta_{\rm CP}$ and $\sin^2\theta_{23}$ parameter space for NO$\nu$A (T2K) in blue (pink) with $90\%$ of C.L., for NO and also the combined analysis in black lines. \ac{It should be noticed that, by combining both experiments, the allowed region is closer to the scenario of T2K only.}
On the left-hand side we show the standard oscillation scenario.
In the middle panel, we have the effects of \ac{CC-}NSI considering only the parameter $\epsilon_{e\mu}$ and in the right-hand side only $\epsilon_{\mu e}$.
In both cases, the regions overlap completely for NO with 90\% of C.L., alleviating the tension between the experiments. \ac{We quantify in the Supplemental Material that the tension decreases in these scenarios.}
In fact, our analyzes were quantified using the GLoBES software~\cite{Huber:2004ka,Huber:2007ji}, whose results are summarized in Table~(\ref{tab:analysis}).

\begin{figure}[!htp]
    \centering
    \hspace{-0.4cm}
    \includegraphics[width=0.245\textwidth]{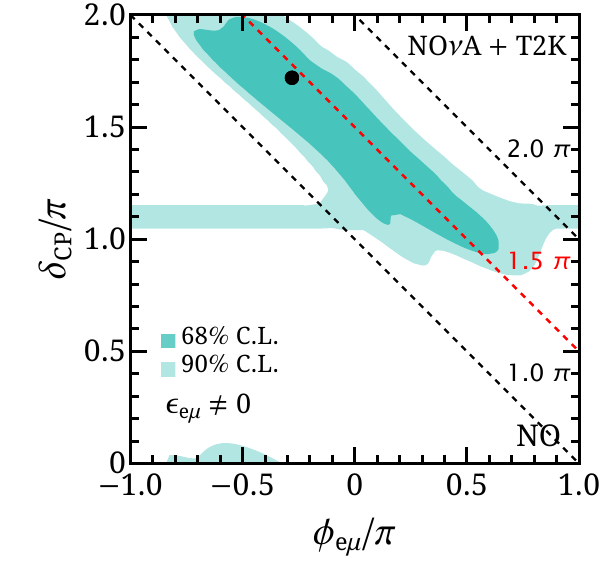}
    \includegraphics[width=0.245\textwidth]{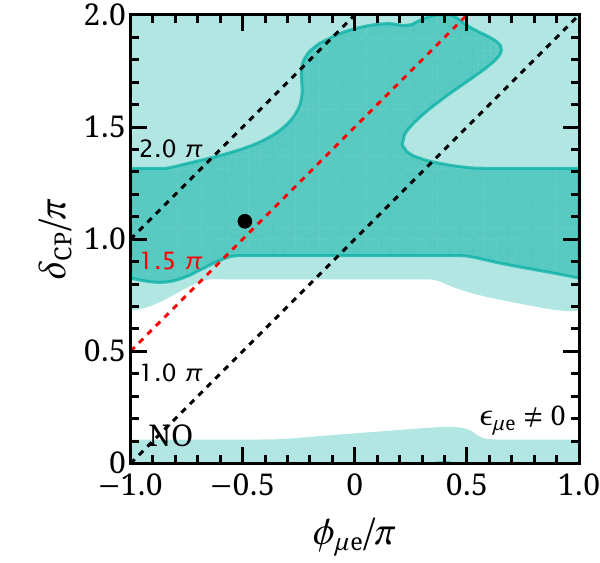}
    \caption{Allowed region for 68 and 90 \% C.L. for combined T2K and NO$\nu$A data. in the parameter space of CP phase of PMNS matrix $\delta_{\rm CP}$ versus the CP phase of \ac{CC-}NSI parameter $\phi_{\alpha \beta}$, $\alpha,\beta=e,\mu$,     for neutrino oscillation with \ac{CC-}NSI. 
    The left (right) panel is for the normal ordering for the scenario of  $\epsilon_{e\mu}$ ( $\epsilon_{\mu e}$ ).     The dot point denotes the best-fit parameters \ac{(not considering the pion decay bounds)}, and the dashed lines indicates specific values for the phases sum, $    \delta_{\rm CP}    +\phi_{e \mu}$     (left) and phases difference, $     \delta_{\rm CP}     -\phi_{\mu e}
     $ (right).
     }
\label{fig:combined_eps_phi}
\end{figure}

\begin{figure}[!htp]
    \centering
    \hspace{-0.4cm}
    \includegraphics[scale=0.4]{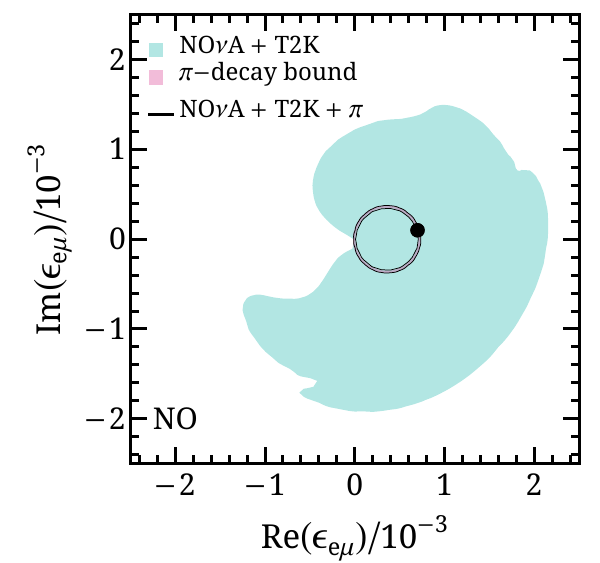}
    \includegraphics[scale=0.4]{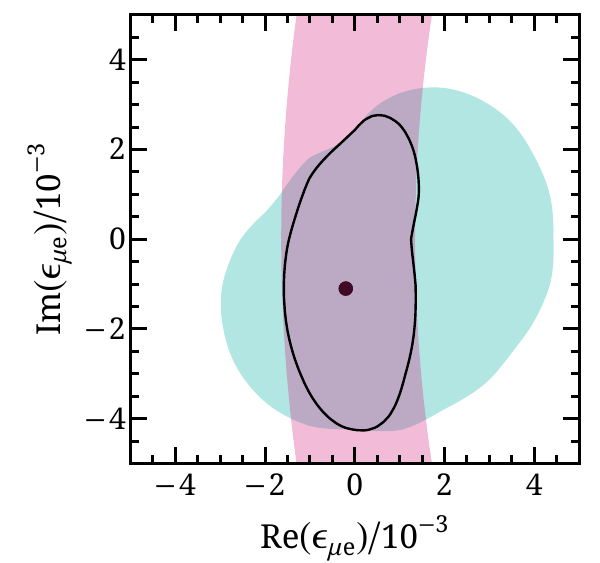} 
    \caption{Bounds for $\epsilon_{e\mu
    }$ ($\epsilon_{\mu e}$) in the left (right) panel, coming from our analysis \ac{(blue curve)}, the constraints from $\pi$-decay rate (pink curve), and considering both bounds - from oscillation and $\pi$-decay experiments - (black curve) at 90\% C.L. The legends are the same for both panels. The best-fit values for the combination are represented by dots.
    }
     \label{fig:2D_combined_V2}
\end{figure}

A fair estimate of the compatibility of a given model for different data sets is given by the goodness-of-fit parameter~\cite{Maltoni:2003cu,Maltoni:2002xd,Machado:2013xiy}. The parameter goodness of fit (PG) is defined as $\chi^2_{\rm PG}\equiv 
\chi^2_\mathrm{min}-\sum_k (\chi^2_k)_\mathrm{min}$, where  $\chi^2_\mathrm{min}$ and  $(\chi^2_k)_\mathrm{min}$ are the global minimum and the local minimum.  
It is illustrative to notice the p-values of the different scenarios in Table~(\ref{tab:analysis}). \ac{As higher the p-value is, more compatible the two experiments are between each other.}
\ac{If the \ac{CC-}NSI contribution is absent, the p-value for NO is $14\%$, while for IO it is $70\%$. It clearly shows the nature of the present tension between the T2K and NO$\nu$A experiments, as each of them, individually, prefers NO. 
By including the \ac{CC-}NSI parameter, the p-value for the NO case increases. Notice that for the $\epsilon_{e\mu}$ scenario, the minimum value for $\chi^{2}_{\rm min}$ occurs for NO for each experiment individually, as well as for their combination. Then it implies that the tension for the $\epsilon_{e\mu}$ scenario is lifted. For the $\epsilon_{\mu e}$ scenario, the tension is reduced compared to the standard oscillation case.}

As seen in Table~(\ref{tab:BF}), the best-fit for the combined analysis \ac{for NO} has $\delta_{\rm CP}$ \ac{and the \ac{CC-}NSI phase} different than zero. It is then natural to ask how sensitive the experiments are to claim that CP is violated in the leptonic sector. In Figure~\ref{fig:combined_eps_phi}, we show the allowed regions with 68 and 90 \% C.L. in the parameter space of phases for \ac{NO}. The left (right) panel corresponds to the parameter space $\delta_{\rm CP} \ {\rm vs.}\ \phi_{e\mu}$ ($\delta_{\rm CP} \ {\rm vs.}\ \phi_{\mu e}$). As anticipated from Eq.~\eqref{eq:prob_dominante}, for $|\epsilon_{\mu e}|\sim \mathcal{O}(10^{-3})$ the conversion probability has a dependence on the phase difference $\delta_{\rm CP}-\phi_{\mu e}$, which explains the tendency seen on right panel of Figure~\ref{fig:combined_eps_phi}. 
For $\epsilon_{e \mu}$, the left panel of Figure~\ref{fig:combined_eps_phi}, there is a dependence on the sum of phases, which is now much more evident. 
Although the best-fit for $\phi_{X}$, $\delta_{\rm CP}$, $\delta_{\rm CP}\pm \phi_{X}$ are all different than 0 (or 2$\pi$), the present data combining T2K and NO$\nu$A is not enough to claim leptonic CP violation in the presence of \ac{CC-}NSI at 1$\sigma$ C.L. or higher.

Finally, we contrast the parameter region allowed by NO$\nu$A and T2K data against constraints from other experiments. The same Lagrangian shown in 
Eq.~(\ref{eq:lagrangian}) can induce changes in the leptonic decay rate of the pion, which is one of the best-measured \ac{observables}~\cite{ParticleDataGroup:2024cfk}. 
We show in Figure~\ref{fig:2D_combined_V2} the allowed region in the real vs. imaginary part of the \ac{CC-}NSI parameter space, for $\epsilon_{e\mu}$ on the left and $\epsilon_{\mu e}$ in the right panel, in blue.
We also show in pink the region allowed by the constraints on pion decay, 
 which is the process with the most stringent bounds to our scenario with \ac{CC-}NSI~\cite{Guzzo:2023ayo,codedecayrate}. The region allowed for neutrino experiments alone is dramatically reduced for the case $\epsilon_{e\mu}$. For the case $\epsilon_{\mu e}$, the main effect is to constrain the real part of the \ac{CC-}NSI parameter. Including data from the neutrino experiments reduces the allowed region in the imaginary axis from the pion decay experiments alone. The previous constraints obtained from neutrino oscillation experiments were $|\epsilon_{\mu e}|< 4\times 10^{-3}$~\cite{Du:2020dwr,Du:2021rdg}  and $|\epsilon_{\mu e}|< 2.6\times 10^{-3}$~\cite{Falkowski:2021bkq} and our limits are more stringent.
 \op{Regarding other proposals in the literature, the overall fit improvement is in general of order of $1\sigma$, but it is not available in general the degree of mitigation of tension. In our scenario,  we have a twofold improvement. We not only have a better overall fit, but also the tension reduces in the CC-NSI scenario.} \ac{In particular, while the standard scenario has a $1.6\sigma$ tension between NO$\nu$A and T2K experiments, for the CC-NSI scenario the tension is reduced to less than $1\sigma$. Moreover, the  $\epsilon_{e\mu}$ scenario has an overall fit improvement of $1.6\sigma$, when compared to the standard scenario. It makes our solution one of the most promising in the literature.}  \op{See the supplemental material~\cite{sup} for more details.}
 
\textit{Discussion \& Conclusion}.--- 
Neutrino oscillation is a unique probe for BSM interactions.
Long-baseline neutrino oscillation experiments are particularly sensitive to non-standard neutrino interaction (\ac{CC-}NSI).
We showed that a new pseudo-scalar four-fermion interaction between quarks and leptons modifies neutrino production. In this scenario, there is a new source of CP violation from the complex \ac{CC-}NSI parameter, $\epsilon_{\mu e}$ or $\epsilon_{e\mu}$
\ac{, thus affecting,} \ac{in particular,} the T2K and NO$\nu$A analyses. We have found for the {\it first time} an analytical formula for neutrino propagation in matter \ac{in the presence of CC-NSI} that is \ac{in excellent} agreement with the numerical solution.

In the literature, the solution for the T2K-NO$\nu$A tension using CC-NSI was not considered before, mostly because there are stringent bounds coming directly from pion decay experiments. We showed that this bias is not justified, by explicitly considering the pion decay bounds as priors in our statistical analyses. On doing so, we have not only reduced the T2K-NO$\nu$A tension, but also provided a better overall fit, in comparison to the standard scenario. The scenario with $\epsilon_{e\mu}$ is particularly promising, being one the best solutions in literature. The non-zero value of the \ac{CC-}NSI parameter opens a new window to understand the source of CP violation, and it can be tested in future neutrino oscillation experiments.

\vspace{0.5cm}
\begin{acknowledgments}
A.C.~acknowledges support from National Council for Scientific and Technological Development – CNPq through projects 166523\slash2020-8 and 201013\slash2022-3.
P.S.P. acknowledges support by the National Natural Science Foundation of the China (12375101, 12090060 and 12090064), the SJTU Double First Class start-up fund (WF220442604) and . the Grant-in-Aid for Innovative Areas No. 19H05810. O.G.L.P. acknowledges support for the FAPESP funding Grant 2014\slash19164-6,2022\slash08954-2, 2021/13757-9 and 2024/07128-7, and the National Council for Scientific and Technological Development – CNPq grant 306565\slash2019-6 and 306405\slash2022-9. P.S.P and O.L.G.P acknowledge support from FAEPEX/UNICAMP 2404/25. This study was financed in part by the Coordenação de Aperfeiçoamento de Pessoal de Nível Superior - Brasil (CAPES) - Finance Code 001.   E. S. S.~acknowledges support from National Council for Scientific and Technological Development - CNPq through Project 140484\slash2023-0.
\end{acknowledgments}

\bibliographystyle{apsrev4-1}
\bibliography{doubly.bib}

\newpage
\appendix

\onecolumngrid

\ifx \standalonesupplemental\undefined
\setcounter{page}{1}
\setcounter{figure}{0}
\setcounter{table}{0}
\setcounter{equation}{0}
\fi

\newpage
\begin{center}
{\Large\bf\boldmath Supplemental Material}
\end{center}	
\vspace{1cm}

This Supplemental material contains a more detailed description of the analytical probabilities derived in this work as well as their comparison with the numerical method. We provide results not considering the pion decay constraint as prior as well as a more detailed comparison to other BSM scenarios considered previously in the literature. We also describe in more detail our statistical analysis.

\section{ Probabilities with NSI at source}

The transition amplitude 
in the presence of a pseudo-scalar interaction $\sqrt{2}\, G_{F} V_{ud}^{\rm CKM} \epsilon_{\alpha \beta} \left(\bar{u} \gamma^5 d \right)\left(\bar{\ell}_{\alpha}P_L \nu_{\beta} \right)$ changes the 
standard oscillation amplitude $S_{\beta\alpha}^{\rm OSC}\equiv \langle \nu_\beta | e^{-i H L} |\nu_\alpha \rangle$ by combining it with the $\epsilon$ matrix,
$S_{\beta \alpha}^{\rm OSC}
\rightarrow 
S_{\beta \alpha}^{\rm NSI} 
= 
(
\delta_{\alpha\alpha'} - p_\alpha \epsilon_{\alpha\alpha'}^*
) 
S_{\beta \alpha' }^{\rm OSC}$. 
Then, the oscillation probability $P_{\alpha \beta}^{\rm NSI} \equiv |S_{\beta \alpha}^{\rm NSI}|^2$ for the two 
transitions become
\begin{align}
P_{\mu e}^{\rm NSI}
    =
    \left|
        S_{e \mu}^{\rm OSC} 
        - 
        p_\mu 
        \epsilon_{\mu e}^{*}
        S_{e e}^{\rm OSC}
    \right|^2,\\
P_{e e}^{\rm NSI}
    =
    \left|
        S_{e e}^{\rm OSC} 
        - 
        p_e 
        \epsilon_{e \mu}^{*}
        S_{e \mu}^{\rm OSC}
    \right|^2 ~ .
\label{eq:Probs_NSI}
\end{align}
respectively for $\epsilon_{\mu e}$ and $\epsilon_{e \mu}$ scenario.
We write the parameters of the NSI complex as
$\epsilon_{\mu e} \equiv |\epsilon_{\mu e}| e^{i\phi_{\mu e}} $ and
$\epsilon_{e\mu} = |\epsilon_{e\mu}|e^{i\phi_{e\mu}}$. The amplitudes
$ S_{e \mu}^{\rm OSC}$ and $ S_{e e}^{\rm OSC}$ were obtained analytically in \cite{Asano:2011nj}
\footnote{Since $S_{e e}^{\rm OSC}$ and $S_{e \mu}^{\rm OSC}$ enter 
in both oscillation probabilities, we can derive a transformation $F$,
acting on the elements of $\epsilon$ matrix in such a way that $P_{\mu e}^{\rm NSI} \xrightarrow{F} P_{e e}^{\rm NSI}$,
\begin{align}
  F
\ : \
    p_\mu \epsilon_{\mu e}^{*} 
    \to
    \frac{
            1
        }
        {
            p_e \epsilon_{e \mu}^{*} 
        } ~ ,
\label{eq:trans}
\end{align}
by the relationship
\begin{align}
    P_{e e}^{\rm NSI}
=
  p_e^2 |\epsilon_{e \mu}^{*}|^2
  \:
  P_{\mu e}^{\rm NSI}
  \!
  \left(
    \left[
        F : p_\mu \epsilon_{\mu e}^{*}
    \right]
  \right) ~.
\label{eq:trans_Prob}
\end{align}}.

We also obtain the perturbative formulas for the oscillation probabilities for the $\nu_{\mu}\rightarrow\nu_{e}$ transition $P_{\mu e}^{\rm NSI} \ (\epsilon_{\mu e}\neq 0)$ and survival $P_{e e}^{\rm NSI} \ (\epsilon_{e\mu}\neq 0)$, with matter effects included for the NSI at source scenario.

The expansion uses the established hierarchy between the oscillation parameters \cite{Asano:2011nj},
\begin{equation}
  r_{\Delta}
~\equiv~
  \Delta m_{21}^2/\Delta m_{31}^2 
  \simeq \zeta ~ ,
\quad
{\rm and}
\quad
  \sin \theta_{13} 
\simeq 
  \sqrt{\zeta} ~ ,
\end{equation}
where $\zeta \sim 0.01$ is the perturbative expansion parameter.

The advantages of developing a perturbative method lie primarily in separating the different orders of contribution \cite{Asano:2011nj}.
This separation allows us to describe analytical solutions and understand which terms are predominant.
Then we can write the evolution matrix and therefore the neutrino transition matrix as a series expansion in powers of $\zeta$,
\begin{align}
S_{\beta\alpha}=[S_{\beta\alpha}]^{(0)}+[S_{\beta\alpha}]^{(1/2)}+[S_{\beta\alpha}]^{(1)}+[S_{\beta\alpha}]^{(3/2)}+\cdots
\label{eqq}
\end{align}
where $[S_{\beta\alpha}]^{(r)}$ with the index $r=0$, $1/2$, 
$1$, $3/2$ denotes the power law dependency of $\zeta$ in the expansion.

The expansion Eq.~(\ref{eqq}) translates in an expansion in  
$P_{\alpha \beta}$. The explicit expression we obtain for the probability is 
%
\begin{align}
\label{as1}
 & 
 P_{\mu e}^{\rm NSI}= \left[P_{\mu e}^{\rm NSI}\right]^{(0)} +
  \left[P_{\mu e}^{\rm NSI}\right]^{(1/2)}
  + \left[P_{\mu e}^{\rm NSI}\right]^{(1)} +
   \left[P_{\mu e}^{\rm NSI}\right]^{(3/2)} 
   + \cdots \\
 &
    \left[P_{\mu e}^{\rm NSI}\right]^{(0)}
= 
    p_\mu^{2} |\epsilon_{\mu e}|^{2} ~ ,
    \label{eq:Tordem0}
    \\
&
    \left[P_{\mu e}^{\rm NSI}\right]^{(1/2)}
= 
    4 p_\mu|\epsilon_{\mu e}|\frac{s_{13}s_{23}
    }{1-r_{a}}\sin{\left(\frac{(1-r_{a})\Delta L}{2}\right)}
    \sin{\left(\delta_{\rm CP}- \phi_{\mu e} + \frac{(1-r_{a})\Delta L}{2}\right)} ~ ,
    \label{eq:Tordem1/2}
    \\
&
    \left[P_{\mu e}^{\rm NSI}\right]^{(1)}
=
    4 (s_{23}^{2}-p_\mu^{2}|\epsilon_{\mu e}|^{2}) \frac{  s^2_{13} }{ (1 - r_{a})^2 }\sin^2 \left(\frac{ (1 - r_{a}) \Delta L }{ 2 }\right) \nonumber
    \\
&
    \hspace{1.8 cm}
    -4 p_\mu|\epsilon_{\mu e}| \frac{s_{12}c_{12}c_{23}r_{\Delta}}{r_{a}}\sin{\left(\frac{r_{a}\Delta L}{2}\right)}\sin{\left( \frac{r_{a}\Delta L }{2}+\phi_{\mu e}\right)} ~ ,
     \label{eq:Tordem1}    \\
&
  \left[P_{\mu e}^{\rm NSI}\right]^{(3/2)}
= 
    8 J_{r} \frac{ r_{\Delta} }{ r_{a} (1 - r_{a}) }
    \cos \left(\delta_{\rm CP}+\frac{ \Delta L }{ 2 } \right)
    \sin \left( \frac{ r_{a} \Delta L }{ 2 }\right)
    \sin \left(\frac{ (1 - r_{a}) \Delta L }{ 2 }\right)\nonumber
    \\
&
    \hspace{1.8 cm}
    +p_\mu|\epsilon_{\mu e}|
    \frac{s_{13}s_{23}}{(1-r_{a})^{3}}
    \Bigg[
    2(1 - r_{a})\Delta L
    \sin{(\delta_{\rm CP}- \phi_{\mu e} + (1-r_{a})\Delta L )}
    \left( 
    2 r_{a} s_{13}^2
    -(1 - r_{a}) r_{\Delta} s_{12}^2
    \right)    
    \nonumber
    \\
 &
    \hspace{1.8 cm}
    +\left[
        -\cos (\delta_{\rm CP} -\phi_{\mu e} )
        +\cos{(
            \delta_{\rm CP} -\phi_{\mu e}
            +(1-r_{a})\Delta L
            )}
    \right] 
    \left[
        (3 + r_{a} (2 + r_{a})) s_{13}^2
        - 2(1 - r_{a}) r_{a} r_{\Delta} s_{12}^2
    \right] 
    \nonumber 
    \\
&
    \hspace{1.8 cm}
    +2 s_{13}^2 
    \left[
        \cos{
            (
            \delta_{\rm CP} - \phi_{\mu e}
            -(1-r_{a}) \Delta L
            )
            }
        -
        \cos{
            (\delta_{\rm CP}-\phi_{\mu e})
            }
    \right]
    \Bigg] ~ ,
   \label{eq:Tordem32}
\end{align}
%
and 
\begin{align}
& 
P_{e e}^{\rm NSI}= \left[P_{e e}^{\rm NSI}\right]^{(0)} +
  \left[P_{e e}^{\rm NSI}\right]^{(1/2)}
  + \left[P_{e e}^{\rm NSI}\right]^{(1)} +
   \left[P_{e e}^{\rm NSI}\right]^{(3/2)}  
   + \cdots   
\\
&
    \left[P_{e e}^{\rm NSI}\right]^{(0)}
    = 
    1 ~ ,
\label{eq:Tordee0}
\\
&
    \left[P_{e e}^{\rm NSI}\right]^{(1/2)}
    =
    4p_e|\epsilon_{e\mu}|\frac{s_{13}s_{23}}{1-r_{a}}\sin{\left(\frac{(1-r_{a})\Delta L}{2}\right)}
    \sin{
        \left(
            \delta_{\rm CP}+\phi_{e\mu}    
            +
            \frac{(1-r_{a})\Delta L}{2}
        \right)
        } ~ ,
\label{eq:Tordee1/2}
\\
&    
    \left[P_{e e}^{\rm NSI}\right]^{(1)}
    = 
    -4 (1-p_e^{2}|\epsilon_{e \mu}|^{2}s_{23}^{2}) \frac{  s^2_{13} }{ (1 - r_{a})^2 }
    \sin^2{{\left(\frac{(1-r_{a})\Delta L}{2}\right)}}\nonumber
    \\
    &
    \hspace{1.8 cm}
    -4p_e|\epsilon_{e\mu}|\frac{s_{12}c_{12}c_{23}r_\Delta}{r_{a}}\sin{\left(\frac{r_{a}\Delta L}{2}\right)}\sin{\left(\frac{r_{a}\Delta L}{2} -\phi_{e\mu}\right)} ~ ,
\nonumber 
\\
&  
  \left[P_{e e}^{\rm NSI}\right]^{(3/2)}
    =
    8 J_r p_e^2 |\epsilon_{e\mu}|^2
    \frac{ r_\Delta }{ (1-r_{a})r_{a} }
    \cos{
        \left(
            \frac{\Delta L}{2}+\delta_{\rm CP} 
        \right)
        }
    \sin{
        \left(
            \frac{ r_{a} \Delta L }{2}
        \right)
        }
    \sin{
        \left(
            \frac{ (1-r_{a}) \Delta L}{2}
        \right)
        }
    \\
    &
    \hspace{1.8 cm}  
    + p_e |\epsilon_{e\mu}|
    \frac{s_{23}s_{13}}{(1-r_{a})^3}
    \bigg[
        2(1-r_{a})\Delta L
        \sin{
            (
                \delta_{\rm CP} + \phi_{e\mu} + (1-r_{a})\Delta L 
            )
            }
        \left(
            2r_a s_{13}^2 
            - (1-r_a)r_\Delta s_{12}^2
        \right)
        \nonumber
        \\
        &
        \hspace{1.8 cm}     
        +
        \left[
            -
            \cos{
                (
                    \delta_{\rm CP}+\phi_{e\mu}
                )
                }
            +
            \cos{
                (
                    \delta_{\rm CP}+\phi_{e\mu}+(1-r_a)\Delta L
                 )
                }
        \right]
        \left[
            (3+r_a(2+r_a))s_{13}^2
            -2(1-r_a)r_a r_\Delta s_{12}^2
        \right]
        \nonumber
        \\     
        &
        \hspace{1.8 cm}
        + 2 s_{13}^2 
        \left[
            \cos{
            (
                \delta_{\rm CP} + \phi_{e\mu} - (1-r_{a}) \Delta L
            )
            }
            -
            \cos{
                (
                    \delta_{\rm CP} + \phi_{e \mu}
                )
                }
    \right]
    \bigg] ~ ,
\label{eq:Tordee32}
\end{align}
where 
$\Delta = \Delta m_{31}^{2}/2E_{\nu}$, 
$L$ is the distance between the source and detector, 
$r_{a} = a/\Delta$ 
with $a=\sqrt{2}G_F N_e$ being the matter potential
and the Jarskolg factor \cite{Jarlskog:1985ht,Jarlskog:1985cw}
$J_{r} = c_{12}s_{12}c_{23}s_{23}s_{13}$ and the cosine and sine of the mixing angles are given in shorthand notation
$s_{ij}=\sin{\theta_{ij}}$ and $c_{ij}=\cos{\theta_{ij}}$.
The oscillation probability for antineutrinos, is obtained by performing the replacements
$
\delta_{CP} \rightarrow -\, \delta_{CP}
$, 
$
\phi_{\mu e} \rightarrow -\, \phi_{\mu e}
$, 
$
\phi_{e \mu} \rightarrow -\, \phi_{e \mu}
$ 
and 
$
a \rightarrow -\, a ~ $.

We highlight a few points about the NSI probabilities
\begin{enumerate}
\item In the usual neutrino oscillation probability the CP violating phase $\delta_{\rm CP}$ appears in lowest order of perturbation as 
$\left[P_{\mu e}^{\rm OSC}\right]^{(3/2)}$.
For the NSI scenario, the CP violating term appears in 
$\left[P_{\alpha e}^{\rm NSI}\right]^{(1/2)}$ 
for 
$\alpha = e, \mu$, 
as can be seen in Eq.~(\ref{eq:Tordem1/2}) and Eq.~(\ref{eq:Tordee1/2});
\item The lowest term that has CP violation effects is given by the combination of the phases
$\delta_{\rm CP}- \phi_{\mu e}$
and
$\delta_{\rm CP}+ \phi_{e \mu}$,
respectively  to  Eq.(\ref{eq:Tordem1/2}) and Eq.(\ref{eq:Tordee1/2}). This behavior is apparent in Figure 4 of the main paper, where the allowed region follows this dependence;
\item The survival probability in the NSI scenario depends on CP phase of the PMNS matrix, $\delta_{\rm CP}$, when in the standard neutrino oscillation scenario the survival probability is independent of this parameter.
\end{enumerate}

\section{Comparison between analytical and exact formula}

In order to ensure the validity of the perturbative formulas derived in the article, we cross-checked the analytical expressions against numerical results using GLoBES~\cite{Huber:2004ka,Huber:2007ji}.
Therefore, in the case of non-zero $\epsilon_{\mu e}$, the $\nu_{\mu}\rightarrow\nu_{e}$ transition is given by Eqs.~(\ref{eq:Tordem0})-(\ref{eq:Tordem32}).
We define the error as the ratio of the difference of conversion probability between the analytical formula and the numerical computation over the average value of the analytical and numerical probability as 
\begin{equation}
   {\rm Error} \ [\%] = 100 \cdot \left|\,\frac{ \big( P_{\mu e}^{\rm Num} - P_{\mu e}^{{\rm Ana}} \big) }{ \big( P_{\mu e}^{{\rm Num}}+P_{\mu e}^{{\rm Ana}} \big)/2 }\,\right|~,
   \label{eq:Error}
\end{equation}
where $P_{\mu e}^{\rm Num}$ and $P_{\mu e}^{\rm Ana}$ correspond to neutrinos conversion for the numerical and analytical case, respectively. 

\begin{figure}[btp]
\centering
    \includegraphics[scale=0.8]{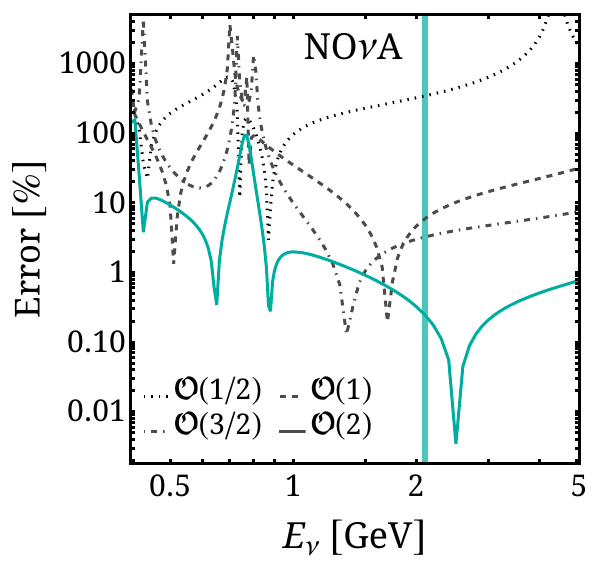}
    \includegraphics[scale=0.8]{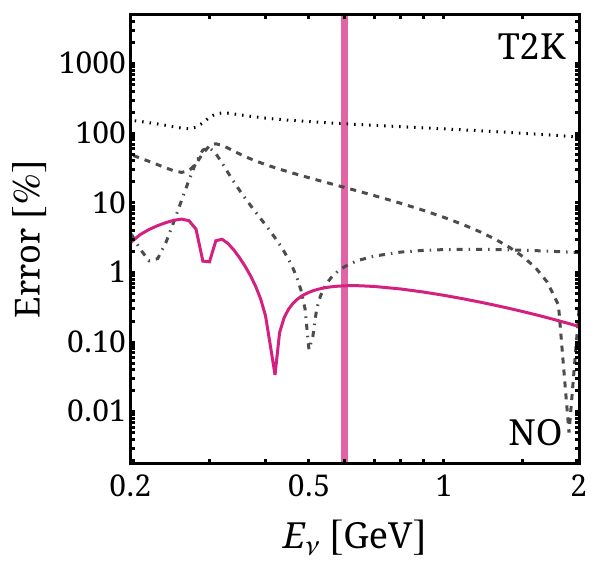}
    \caption{
      Relative error 
      (Eq.~(\ref{eq:Error}))
      between the numerical 
      and analytical transition rate (Eqs.~(\ref{eq:Tordem0})-(\ref{eq:Tordem32}), including second-order) for NO. 
      On the left (right) side, we plot with the best-fit values $\sin^{2}\theta_{23} = 0.57 (0.51)$,
      $\delta_{\rm CP}/\pi = 0.01 (1.44)$,
      $|\epsilon_{\mu e}|/10^{-3} = 1.49 (1.21)$
      and
      $\phi_{\mu e}/\pi = 0.56 (-0.73)$
      for NO$\nu$A (T2K).
      Each curve represents the truncated order up to $\mathcal{O}(1/2)$ (dotted), 
      $\mathcal{O}(1)$ (dashed), $\mathcal{O}(3/2)$ (dot-dashed) 
      and $\mathcal{O}(2)$ (continuous).
      }
    \label{fig:Error_NO_NSI}
\end{figure}

In Supplemental Figure~\ref{fig:Error_NO_NSI}, we show the relative error as given by Eq.~(\ref{eq:Error}) for the conversion rate of neutrinos in the energy range of  NO$\nu$A and T2K experiments for the case of normal ordering (NO) and with NSI. 
As shown in Eq.~(\ref{as1}), the conversion probability can be decomposed in a series of terms, which are plotted as different lines in Supplemental
Figure~\ref{fig:Error_NO_NSI}, $
P_{\mu e}^{\rm NSI} 
\equiv 
 \displaystyle\sum_n \left[P_{\mu e}^{\rm NSI}\right]^{(n)}$ with $n$ truncated to the corresponding order $\mathcal O(n)$. 

Note that the truncated order at $\mathcal{O}(2)$ presents relative errors less than $1\%$ 
at $2.1$~GeV for NO$\nu$A and $0.6$~GeV for T2K (denoted by a vertical line in Supplemental Figure~(\ref{fig:Error_NO_NSI})), that is the typical
value of the energy spectrum of both experiments.
Similar studies were performed for the case of antineutrino conversion probability, resulting in analogous conclusions.

\section{Results using only oscillation data}

In this section we present our results for the NO$\nu$A and T2K simulations without including the pion decay bounds as priors. We begin with the minimum values for $\chi^{2}$ for the standard oscillation, $\epsilon_{e\mu}$ and $\epsilon_{\mu e}$ scenarios in Table~(\ref{tab:analysis}).

\begin{table}[!htp]
\centering
\begin{tabular}{c||c|c||c|c||c|c}
\multirow{2}{*}{$\chi^2_{\rm min}$}
& 
 \multicolumn{2}{c||}{Standard Osc.}
&
 \multicolumn{2}{c||}{$\epsilon_{e\mu}$} 
&
  \multicolumn{2}{c}{$\epsilon_{\mu e}$}
\\ 
  & NO & IO
  & NO & IO
  & NO & IO
\\[1pt] \hline \hline
\noalign{\vskip 1.5pt}
  NO$\nu$A      &  \hspace{0.1 pt} 51.8 \hspace{0.1 pt} 
                &  \hspace{0.1 pt} 52.5 \hspace{0.1 pt} 
                &  \hspace{0.1 pt} 48.4 \hspace{0.1 pt}
                &  \hspace{0.1 pt} 50.4 \hspace{0.1 pt} 
                &  \hspace{0.1 pt} 51.3 \hspace{0.1 pt}
                &  \hspace{0.1 pt} 51.6 \hspace{0.1 pt}
\\
  T2K           &  \hspace{0.1 pt} 107.2                          \hspace{0.1 pt}
                &  \hspace{0.1 pt} 109.2 \hspace{0.1 pt}
                &  \hspace{0.1 pt} 106.3 \hspace{0.1 pt}
                &  \hspace{0.1 pt} 107.6 \hspace{0.1 pt}
                &  \hspace{0.1 pt} 106.5 \hspace{0.1 pt}
                &  \hspace{0.1 pt} 106.8 \hspace{0.1 pt}
\\
NO$\nu$A + T2K \hspace{0.1 pt}
                &  \hspace{0.1 pt} 160.75.9 \hspace{0.1 pt}
                &  \hspace{0.1 pt} 163.9 \hspace{0.1 pt}
                &  \hspace{0.1 pt} 161.4 \hspace{0.1 pt}
                &  \hspace{0.1 pt} 161.0 \hspace{0.1 pt}
                &  \hspace{0.1 pt} 165.2 \hspace{0.1 pt}
                &  \hspace{0.1 pt} 162.4 \hspace{0.1 pt}
\\[1.0pt] \hline \hline
\noalign{\vskip 1.5pt}
$\chi^2_{\rm \textbf{PG}}\textbf{\,/\,N}_{\rm \textbf{par}}$ 
& \hspace{0.1 pt} 7.0\,/\,4 \hspace{0.1 pt}
& \hspace{0.1 pt} 2.2\,/\,4 \hspace{0.1 pt}
& \hspace{0.1 pt} 6.7\,/\,6 \hspace{0.1 pt}
& \hspace{0.1 pt} 3.0\,/\,6 \hspace{0.1 pt}
& \hspace{0.1 pt} 7.4\,/\,6 \hspace{0.1 pt}
& \hspace{0.1 pt} 4.0\,/\,6 \hspace{0.1 pt}
\\ \hline 
\textbf{p}$_{\rm \textbf{PG} }$\textbf{-value} 
&  \hspace{0.1 pt} 14\% \hspace{0.1 pt}
&  \hspace{0.1 pt} 70\% \hspace{0.1 pt}
&  \hspace{0.1 pt} 35\% \hspace{0.1 pt}  
&  \hspace{0.1 pt} 81\% \hspace{0.1 pt}
&  \hspace{0.1 pt} 28\% \hspace{0.1 pt} 
&  \hspace{0.1 pt} 68\% \hspace{0.1 pt}
\end{tabular}
\caption{
We present the results of the standard oscillation model and for the production \ac{CC-}NSI of the values of $\chi^2$ minimum for the individual datasets of  NO$\nu$A and T2K and the combined analysis. The values of PG test are listed for the four free parameters in the standard oscillation scenario and six with \ac{CC-}NSI.
}
\label{tab:analysis}
\end{table}

We also present the best-fit values found for NO for the CC-NSI scenarios without (wiht) pion decay bounds in 
Table~(\ref{tab:BF}).

\begin{table}[!htp]
\centering
\begin{tabular}{c ccc}
\multicolumn{1}{ c |}
    { 
        \textbf{NO} $(\epsilon_{e\mu}\neq 0)$ 
        \hspace{2 pt}
    }
 &  
    \hspace{4 pt} 
        NO$\nu$A (+$\pi$)
    \hspace{2 pt}
 & 
    \hspace{1 pt} 
        T2K (+$\pi$)
    \hspace{2 pt}
 & 
    \hspace{1 pt} 
        NO$\nu$A + T2K (+$\pi$)
 \\[1pt] \hline
\\[-18 pt]
\multicolumn{1}{ c |}{  }
\\
\multicolumn{1}{ c |}
    { 
        $\sin^{2}\theta_{23}/10^{-1}$ 
        \hspace{1 pt}
    }
            &  5.71 (\ac{4.52})& 5.39 (\ac{5.47}) &  5.68 (\ac{5.68})
\\[1pt]
\multicolumn{1}{ c |}
    { 
        $\delta_{\rm CP}/\pi$ 
        \hspace{1 pt}
    } 
            & 0.04 (\ac{1.81}) & 1.28 (\ac{1.39}) & 1.74 (\ac{1.35})
\\[1pt]
\multicolumn{1}{ c |}
    {
        $|\epsilon_{e\mu}|/10^{-3}$
        \hspace{1 pt}
    }
            & 1.29 (\ac{0.54}) & 1.75 (\ac{0.054}) & 0.72 (\ac{0.70})
\\[1pt]
\multicolumn{1}{ c |}
    { 
        $\phi_{e\mu}/\pi$ 
        \hspace{1 pt}
    }
            & -0.80 (-\ac{0.23}) & -0.26 (\ac{0.48}) & -0.32 (\ac{0.08})
\\[1pt]
\multicolumn{1}{ c |}
    { 
        (
        $
        \delta_{\rm CP}
        +
        \phi_{e\mu}
        )
        /\pi
        $
        \hspace{1 pt}
    }
            & 1.24 (\ac{1.58}) & 1.02 (\ac{1.87}) & 1.42 (\ac{1.43})
\\[3 pt] \hline
 \end{tabular}
 \\[11.0 pt]
%
\begin{tabular}{c ccc}
\multicolumn{1}{ c |}
    { 
        \textbf{NO}
        $(\epsilon_{\mu e}\neq 0)$ 
        \hspace{2 pt}
    }
 &  
    \hspace{4 pt} 
        NO$\nu$A (+$\pi$)
    \hspace{2 pt}
 & 
    \hspace{1 pt} 
        T2K (+$\pi$)
    \hspace{2 pt}
 & 
    \hspace{1 pt} 
        NO$\nu$A + T2K (+$\pi$)
\\[1pt] \hline
\\[-18 pt]
\multicolumn{1}{ c |}{  }
\\
\multicolumn{1}{ c |}
    { 
        $\sin^{2}\theta_{23}/10^{-1}$ 
        \hspace{1 pt}
    }
            &  5.73 (\ac{5.72}) & 5.01 (\ac{5.32}) &  5.69 (\ac{5.69})
\\[1pt]
\multicolumn{1}{ c |}
    { 
        $\delta_{\rm CP}/\pi$ 
        \hspace{1 pt}
    } 
            & 0.01 (\ac{1.98}) & 1.32 (\ac{1.34}) & 1.08 (\ac{1.06})
\\[1pt]
\multicolumn{1}{ c |}
    {
        $|\epsilon_{\mu e}|/10^{-3}$
        \hspace{1 pt}
    }
            & 1.52 (\ac{1.63}) & 1.14 (\ac{0.45}) & 1.12 (\ac{1.09})
\\[1pt]
\multicolumn{1}{ c |}
    { 
        $\phi_{\mu e}/\pi$ 
        \hspace{1 pt}
    }
            & 0.57 (\ac{0.53}) & -1.01 (\ac{-0.73}) & -0.49 (\ac{-0.53})
\\[1pt]
\multicolumn{1}{ c |}
    { 
        (
        $
        \delta_{\rm CP}
        -
        \phi_{\mu e}
        )
        /\pi
        $
        \hspace{1 pt}
    }
            & 1.44 (\ac{1.45}) & 0.33 (\ac{0.07}) & 1.57 (\ac{1.59})
\\[2pt] \hline
 \end{tabular}
\caption{The best-fit values for the normal ordering (NO) with the \ac{CC-}NSI $\epsilon_{e\mu}$ in the top table and $\epsilon_{\mu e}$ in the bottom table.
The phases $\delta_{\rm CP}$, $\phi_{e\mu}$, and $\phi_{\mu e}$ are invariant under addition of multiples of $2\pi$.
}
\label{tab:BF}
\end{table}

Finally, in Figure~\ref{fig:2D_combined_V2a} we present the allowed region in the plane $\sin^{2}\theta_{23}$ versus $\delta_{\rm CP}$, if the bound from pion decay is not taken into account.

\begin{figure*}[h!]
    \centering
    \includegraphics[scale=0.55]{FiguresUpdated/delta_sin_SM_NO_Final.pdf}
    \includegraphics[scale=0.55]{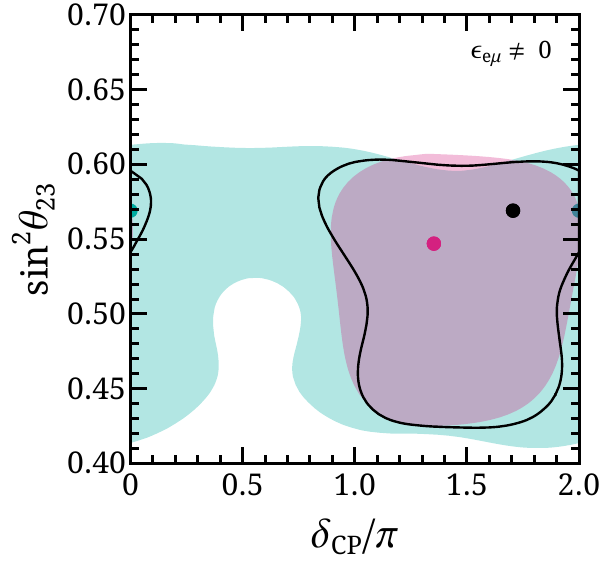}
    \includegraphics[scale=0.55
    ]{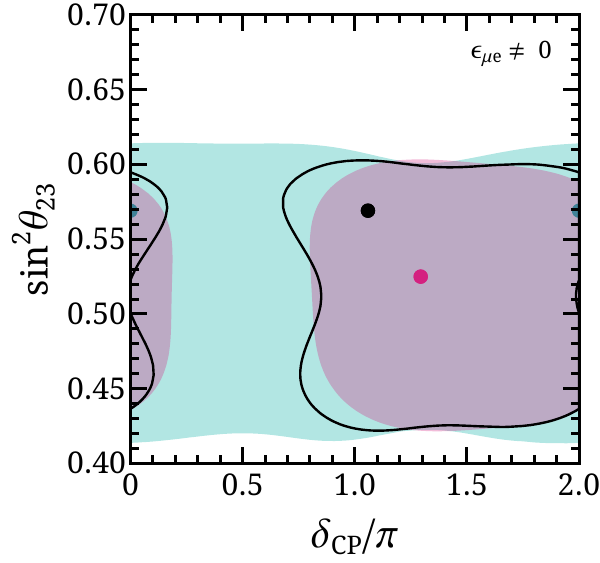}
    \caption{Allowed region for T2K (pink), NOvA (blue) and for combined analysis (black line), for NO in the $\sin^2 \theta_{23}$ vs. $\delta_{\rm CP}$ space, for 90$\%$ confidence level. In the left panel we show the standard oscillation scenario, in the middle panel  
    we show the case with $\epsilon_{e\mu} \neq 0$ and in the right panel the case with $\epsilon_{\mu e} \neq 0$.
    The dots are the respective best-fit values, see Table~(\ref{tab:BF}).
    }
     \label{fig:2D_combined_V2a}
\end{figure*}

\section{Comparison with other BSM scenarios}

The tension between the combination NO$\nu$A and T2K is visible in Figure 3 in the main document. 

We compiled the BSM studies used to address the T2K-NO$\nu$A tension in Table~(\ref{tab:results}) comparing two statistical analysis, the parameter goodness of fit (\textbf{PG})~\cite{Maltoni:2003cu,Maltoni:2002xd,Machado:2013xiy} and the goodness of fit(\textbf{G}). The former measure the internal consistence of the analysis of different experiments into a combined analysis of all experiments. The latter measure the improvement of the quality of the fit between two nested models.

In Table~(\ref{tab:results}) we compile the parameter goodness of fit,
\begin{eqnarray}
    \Delta\chi^2_{\rm \textbf{PG}} ({\rm NO}\nu{\rm A + T2K} )\equiv \chi^2_{{\rm NO} \nu {\rm A + T2K}} 
    -\chi^2_{{\rm NO}\nu{\rm A}}-\chi^2_{{\rm T2K}}
\end{eqnarray}
for a given scenario, standard oscillation or CC-NSI.
The goodness of fit is
\begin{eqnarray}
    \Delta\chi^2_{\rm \textbf{G}} ({\rm NO}\nu {\rm A + T2K} )\equiv \chi^2_{{\rm NO}\nu {\rm A + T2K}} ({\rm CC-NSI})-\chi^2_{{\rm NO}\nu {\rm A + T2K}} ({\rm Standard\, osc.})
\end{eqnarray}
These statistical tests measured the following. internal consistence of a given scenario for the combination of the two experiments and the relative quality of fit of the CC-NSI scenario compared with the standard oscillation scenario. We can translate the number of standard deviations($\sigma$). For our results, listed in the first three lines of Table~(\ref{tab:results}) we have found that {\it in this work}, from the $\Delta\chi^2_{\rm \textbf{PG}} ({\rm NO}\nu{\rm A + T2K} )$ values listed in Table~(\ref{tab:results}), that the standard oscillation solution for the combination is unfavorable by $1.6\sigma$. For both  $\epsilon_{e\mu}$ CC-NSI scenario and $\epsilon_{\mu e}$ CC-NSI scenario, the combined solution is less favorable than $1\sigma$, which implies that the CC-NSI scenario decreases the tension between the NO$\nu$A and T2K. For most of the other analysis, there is no public information available for the $\Delta\chi^2_{\rm \textbf{PG}} ({\rm NO}\nu{\rm A + T2K} )$ test.  

For the quality of fit, we notice that our work improves by 1.6 $\sigma$  over the standard neutrino oscillation scenario as indicated in Table~(\ref{tab:results}). 
All results discussed above are for normal ordering (NO), for the inverted ordering (IO), to include or not include CC-NSI did not change much the quality of fit compared with standard oscillation scenario.

The BSM models that were analyzed with T2K and NO$\nu$A data, respectively, non-unitary mixing matrix\cite{Dutta:2016vcc,Miranda:2019ynh,Yu:2024nkc}  neutral current NSI in propagation,\cite{Denton:2020uda,Chatterjee:2020kkm,Chatterjee:2024kbn}
light~\cite{Chatterjee:2020yak} and very light sterile neutrinos~\cite{deGouvea:2022kma},  Lorentz violation~\cite{Rahaman:2021leu,Rahaman:2022rfp} and dark photon scenario~\cite{Lin:2023xyk,Alonso-Alvarez:2024wnh}.
\begin{table}[!htp]
\centering
\begin{tabular}{c||c|c|c|c|c|c|c|c|c|c|c|c|c|c|c|}
\multirow{2}{*}{Scenario}
& 
 \multicolumn{3}{c|}{ NO$\nu$A}
&
 \multicolumn{3}{c|}{T2K     } 
&
  \multicolumn{4}{c|}{
$\Delta\chi^2_{\rm \textbf{PG}}$ (NO$\nu$A + T2K )}

&  \multicolumn{4}{c|}{$\Delta\chi^2_{\rm \textbf{G}}$ (NO$\nu$A + T2K )}\\ 
  & NO & IO & Par.
  & NO & IO & Par.
  & NO & IO & Par. & $\sigma$ 
    & NO & IO & Par. & $\sigma$ 
    \\  [1pt] \hline \hline
\noalign{\vskip 1.5pt}
Standard Osc 
                &  \hspace{0.1 pt} 51.8 \hspace{0.1 pt}
                &  \hspace{0.1 pt} 52.5 \hspace{0.1 pt}
                &  \hspace{0.1 pt} 4 \hspace{0.1 pt}
                &  \hspace{0.1 pt} 107.2 \hspace{0.1 pt} 
                &  \hspace{0.1 pt} 109.2\hspace{0.1 pt}
                &  \hspace{0.1 pt} 4 \hspace{0.1 pt}
                                &  \hspace{0.1 pt}  6.9\hspace{0.1 pt} 
                &  \hspace{0.1 pt} 2.2 \hspace{0.1 pt}
                &  \hspace{0.1 pt} 4 \hspace{0.1 pt}
                &  \hspace{0.1 pt} 1.6  \hspace{0.1 pt}
                &  \hspace{0.1 pt}  \hspace{0.1 pt}
                &  \hspace{0.1 pt}  \hspace{0.1 pt}
                              &  \hspace{0.1 pt}  \hspace{0.1 pt}
                &  \hspace{0.1 pt}  \hspace{0.1 pt}              
                \\
                CC-NSI $\epsilon_{e\mu}$(This work)\hspace{0.1 pt}                      \hspace{0.1 pt}
                &  \hspace{0.1 pt} 49.3 \hspace{0.1 pt}
                &  \hspace{0.1 pt} 52.1 \hspace{0.1 pt}
                &  \hspace{0.1 pt}6\hspace{0.1 pt}
                &  \hspace{0.1 pt} 107.1 \hspace{0.1 pt}
                &  \hspace{0.1 pt} 108.6 \hspace{0.1 pt}
              &  \hspace{0.1 pt} 6 \hspace{0.1 pt}
                &  \hspace{0.1 pt} 5.4 \hspace{0.1 pt}
                &  \hspace{0.1 pt} 3.2 \hspace{0.1 pt}
                               &  \hspace{0.1 pt} 6 \hspace{0.1 pt}
                                  &  \hspace{0.1 pt}  $< 1$  \hspace{0.1 pt}                &  \hspace{0.1 pt} 4.1 \hspace{0.1 pt}
                            &  \hspace{0.1 pt} 0.0 \hspace{0.1 pt}    
                              &  \hspace{0.1 pt} 2 \hspace{0.1 pt}
              &  \hspace{0.1 pt} 1.6   \hspace{0.1 pt}
                              \\\hspace{0.1 pt}
CC-NSI $\epsilon_{\mu e}$ (This work) &  \hspace{0.1 pt} 51.3 \hspace{0.1 pt}
                &  \hspace{0.1 pt} 51.9 \hspace{0.1 pt}
                &  \hspace{0.1 pt} 6 \hspace{0.1 pt}
                &  \hspace{0.1 pt} 108.6 \hspace{0.1 pt}
                &  \hspace{0.1 pt} 107.0 \hspace{0.1 pt}
                                   &  \hspace{0.1 pt}6 \hspace{0.1 pt}
             &  \hspace{0.1 pt} 4.0 \hspace{0.1 pt}
                &  \hspace{0.1 pt} 3.5  \hspace{0.1 pt}
              &  \hspace{0.1 pt} 6 \hspace{0.1 pt}
               &  \hspace{0.1 pt} $< 1$   \hspace{0.1 pt}                  &  \hspace{0.1 pt} 2 \hspace{0.1 pt}                      &  \hspace{0.1 pt} 1.5  \hspace{0.1 pt}  &  \hspace{0.1 pt} 2  \hspace{0.1 pt} 
                &  \hspace{0.1 pt} $<1$ \hspace{0.1 pt}                \\\hline \hline
                Standard Osc~\cite{Miranda:2019ynh} 
&  \hspace{0.1 pt} 47.9 \hspace{0.1 pt} 
 &  \hspace{0.1 pt} 50.6 \hspace{0.1 pt} 
                &  \hspace{0.1 pt} 4 \hspace{0.1 pt}
                &  \hspace{0.1 pt}  123.7\hspace{0.1 pt} 
                &  \hspace{0.1 pt} 130.7\hspace{0.1 pt}
                &  \hspace{0.1 pt} 4 \hspace{0.1 pt}
                                &  \hspace{0.1 pt}  1.8\hspace{0.1 pt} 
                &  \hspace{0.1 pt} 0.8 \hspace{0.1 pt}
                &  \hspace{0.1 pt} 4 \hspace{0.1 pt}
                &  \hspace{0.1 pt}$<1$  \hspace{0.1 pt}
                &  \hspace{0.1 pt}  \hspace{0.1 pt}
                 &  \hspace{0.1 pt}  \hspace{0.1 pt}
                &  \hspace{0.1 pt}  \hspace{0.1 pt}
                &  \hspace{0.1 pt}  \hspace{0.1 pt}\\
non-unitary~\cite{Miranda:2019ynh} &  \hspace{0.1 pt}44.3\hspace{0.1 pt}
                &  \hspace{0.1 pt} 45.7 \hspace{0.1 pt}
                &  \hspace{0.1 pt} 8 \hspace{0.1 pt}
                &  \hspace{0.1 pt} 121.4 \hspace{0.1 pt}
                &  \hspace{0.1 pt} 123.9 \hspace{0.1 pt}
                &  \hspace{0.1 pt} 8 \hspace{0.1 pt}
                                &  \hspace{0.1 pt} 5.2 \hspace{0.1 pt}
                &  \hspace{0.1 pt} 1.0 \hspace{0.1 pt}
                          &  \hspace{0.1 pt} 8 \hspace{0.1 pt}
                     & \hspace{0.1 pt}    $<1$       \hspace{0.1 pt}                 &  \hspace{0.1 pt} 2.5 \hspace{0.1 pt}
                 &  \hspace{0.1 pt} 5.8 \hspace{0.1 pt}
                &  \hspace{0.1 pt} 4 \hspace{0.1 pt}
 &  \hspace{0.1 pt}$<1$\hspace{0.1 pt}\\[1.0pt] \hline \hline
                NC-NSI $\epsilon_{e\mu}$~\cite{Chatterjee:2020kkm,Chatterjee:2024kbn} \hspace{0.1 pt}                      \hspace{0.1 pt}
                &  \hspace{0.1 pt}  \hspace{0.1 pt}
                &  \hspace{0.1 pt}  \hspace{0.1 pt}
                &  \hspace{0.1 pt}6\hspace{0.1 pt}
                &  \hspace{0.1 pt}  \hspace{0.1 pt}
                &  \hspace{0.1 pt}  \hspace{0.1 pt}
              &  \hspace{0.1 pt} 6 \hspace{0.1 pt}
                &  \hspace{0.1 pt}  \hspace{0.1 pt}
                &  \hspace{0.1 pt}  \hspace{0.1 pt}
                      &  \hspace{0.1 pt}  \hspace{0.1 pt}
                &  \hspace{0.1 pt}  \hspace{0.1 pt}
                      &  \hspace{0.1 pt}4.5 \hspace{0.1 pt}
                &  \hspace{0.1 pt} 0.1 \hspace{0.1 pt}
                &  \hspace{0.1 pt}2\hspace{0.1 pt}
                &  \hspace{0.1 pt} 1.5  \hspace{0.1 pt}
                \\\hspace{0.1 pt}
NC-NSI $\epsilon_{e \tau}$
~\cite{Chatterjee:2020kkm,Chatterjee:2024kbn} &  \hspace{0.1 pt}  \hspace{0.1 pt}
                &  \hspace{0.1 pt}  \hspace{0.1 pt}
                &  \hspace{0.1 pt} 6 \hspace{0.1 pt}
                &  \hspace{0.1 pt}  \hspace{0.1 pt}
                &  \hspace{0.1 pt}  \hspace{0.1 pt}
                &  \hspace{0.1 pt} 6 \hspace{0.1 pt}
                              &  \hspace{0.1 pt}  \hspace{0.1 pt}
                &  \hspace{0.1 pt}  \hspace{0.1 pt}
                           &  \hspace{0.1 pt} \hspace{0.1 pt}
                &  \hspace{0.1 pt}  \hspace{0.1 pt}
                      &  \hspace{0.1 pt}3.8\hspace{0.1 pt}                      &  \hspace{0.1 pt}0\hspace{0.1 pt}
                &  \hspace{0.1 pt} 2 \hspace{0.1 pt}
                      &  \hspace{0.1 pt}1.4 \hspace{0.1 pt}             
                \\[1.0pt] \hline \hline
                NC-NSI $\epsilon_{e\mu}$~\cite{Denton:2020uda}  \hspace{0.1 pt}                      \hspace{0.1 pt}
                &  \hspace{0.1 pt}  \hspace{0.1 pt}
                &  \hspace{0.1 pt}  \hspace{0.1 pt}
                &  \hspace{0.1 pt}6\hspace{0.1 pt}
                &  \hspace{0.1 pt}  \hspace{0.1 pt}
                &  \hspace{0.1 pt}  \hspace{0.1 pt}
              &  \hspace{0.1 pt} 6 \hspace{0.1 pt}
                &  \hspace{0.1 pt}  \hspace{0.1 pt}
                &  \hspace{0.1 pt}  \hspace{0.1 pt}
                      &  \hspace{0.1 pt}  \hspace{0.1 pt}
                &  \hspace{0.1 pt}  \hspace{0.1 pt}
                      &  \hspace{0.1 pt}4.4 \hspace{0.1 pt}
                &  \hspace{0.1 pt} 0.2 \hspace{0.1 pt}
                &  \hspace{0.1 pt}2\hspace{0.1 pt}
                &  \hspace{0.1 pt} 1.5  \hspace{0.1 pt}
                \\\hspace{0.1 pt}
NC-NSI $\epsilon_{e \tau}$~\cite{Denton:2020uda} &  \hspace{0.1 pt}  \hspace{0.1 pt}
                &  \hspace{0.1 pt}  \hspace{0.1 pt}
                &  \hspace{0.1 pt} 6 \hspace{0.1 pt}
                &  \hspace{0.1 pt}  \hspace{0.1 pt}
                &  \hspace{0.1 pt}  \hspace{0.1 pt}
                &  \hspace{0.1 pt} 6 \hspace{0.1 pt}
                              &  \hspace{0.1 pt}  \hspace{0.1 pt}
                &  \hspace{0.1 pt}  \hspace{0.1 pt}
                           &  \hspace{0.1 pt} \hspace{0.1 pt}
                &  \hspace{0.1 pt}  \hspace{0.1 pt}
                      &  \hspace{0.1 pt}3.7\hspace{0.1 pt}                      &  \hspace{0.1 pt}0.7\hspace{0.1 pt}
                &  \hspace{0.1 pt} 2 \hspace{0.1 pt}
                      &  \hspace{0.1 pt}1.4 \hspace{0.1 pt}             
                \\
                [1.0pt] \hline \hline
Sterile neutrino~\cite{Chatterjee:2020yak}&  \hspace{0.1 pt}  \hspace{0.1 pt}
                &  \hspace{0.1 pt}  \hspace{0.1 pt}
                &  \hspace{0.1 pt} 6 \hspace{0.1 pt}
                &  \hspace{0.1 pt}  \hspace{0.1 pt}
                &  \hspace{0.1 pt}  \hspace{0.1 pt}
                &  \hspace{0.1 pt} 6 \hspace{0.1 pt}
                              &  \hspace{0.1 pt}  \hspace{0.1 pt}
                &  \hspace{0.1 pt}  \hspace{0.1 pt}
                           &  \hspace{0.1 pt} \hspace{0.1 pt}
                &  \hspace{0.1 pt}  \hspace{0.1 pt}
                      &  \hspace{0.1 pt}0.7\hspace{0.1 pt}                      &  \hspace{0.1 pt}0.9\hspace{0.1 pt}
                &  \hspace{0.1 pt} 2 \hspace{0.1 pt}
                      &  \hspace{0.1 pt}$<1$ \hspace{0.1 pt}             
                \\[1.0pt] \hline \hline
                Standard Osc ~\cite{Rahaman:2021leu}                &  \hspace{0.1 pt}\hspace{0.1 pt}
                &  \hspace{0.1 pt} \hspace{0.1 pt}
                &  \hspace{0.1 pt} 4 \hspace{0.1 pt}
                &  \hspace{0.1 pt}  \hspace{0.1 pt} 
                &  \hspace{0.1 pt} \hspace{0.1 pt}
                &  \hspace{0.1 pt} 4 \hspace{0.1 pt}
                                &  \hspace{0.1 pt} 2.6 \hspace{0.1 pt} 
                &  \hspace{0.1 pt}  \hspace{0.1 pt}
                &  \hspace{0.1 pt} 4\hspace{0.1 pt}
                      &  \hspace{0.1 pt} $<1$ \hspace{0.1 pt}
                &  \hspace{0.1 pt}  \hspace{0.1 pt}
                      &  \hspace{0.1 pt}  \hspace{0.1 pt}
                &  \hspace{0.1 pt}  \hspace{0.1 pt}
                      &  \hspace{0.1 pt}  \hspace{0.1 pt}
                             \\
Lorentz violation~\cite{Rahaman:2021leu}&  \hspace{0.1 pt}  \hspace{0.1 pt}
                &  \hspace{0.1 pt}  \hspace{0.1 pt}
                &  \hspace{0.1 pt} 7 \hspace{0.1 pt}
                &  \hspace{0.1 pt}  \hspace{0.1 pt}
                &  \hspace{0.1 pt}  \hspace{0.1 pt}
                &  \hspace{0.1 pt} 7\hspace{0.1 pt}
                              &  \hspace{0.1 pt} 4.2 \hspace{0.1 pt}
                &  \hspace{0.1 pt}  \hspace{0.1 pt}
                           &  \hspace{0.1 pt}7 \hspace{0.1 pt}
                &  \hspace{0.1 pt} 1.1 \hspace{0.1 pt}
                      &  \hspace{0.1 pt}2.1\hspace{0.1 pt}  
                      &  \hspace{0.1 pt}\hspace{0.1 pt}
                &  \hspace{0.1 pt} 3 \hspace{0.1 pt}
                      &  \hspace{0.1 pt}$<1$\hspace{0.1 pt}             
                \\[1.0pt] \hline \hline       
Very light Sterile neutrino~\cite{deGouvea:2022kma}&  \hspace{0.1 pt}  \hspace{0.1 pt}
                &  \hspace{0.1 pt}  \hspace{0.1 pt}
                &  \hspace{0.1 pt} 6 \hspace{0.1 pt}
                &  \hspace{0.1 pt}  \hspace{0.1 pt}
                &  \hspace{0.1 pt}  \hspace{0.1 pt}
                &  \hspace{0.1 pt} 6 \hspace{0.1 pt}
                              &  \hspace{0.1 pt} 2.8 \hspace{0.1 pt}
                &  \hspace{0.1 pt}   \hspace{0.1 pt}
                           &  \hspace{0.1 pt} \hspace{0.1 pt}
                &  \hspace{0.1 pt}$<1$\hspace{0.1 pt}
                      &  \hspace{0.1 pt}5.0\hspace{0.1 pt}                      &  \hspace{0.1 pt}\hspace{0.1 pt}
                &  \hspace{0.1 pt} 2 \hspace{0.1 pt}
                      &  \hspace{0.1 pt} $<1$\footnote{This result is listed in   Ref.~\cite{deGouvea:2022kma}.} \hspace{0.1 pt}             
                \\[1.0pt] \hline \hline  
$L_e-L_\mu$ pol.~\cite{Lin:2023xyk}&  \hspace{0.1 pt}  \hspace{0.1 pt}
                &  \hspace{0.1 pt}  \hspace{0.1 pt}
                &  \hspace{0.1 pt} 6\hspace{0.1 pt}
                &  \hspace{0.1 pt} \hspace{0.1 pt}
                &  \hspace{0.1 pt}  \hspace{0.1 pt}
                &  \hspace{0.1 pt}6 \hspace{0.1 pt}
                              &  \hspace{0.1 pt}  \hspace{0.1 pt}
                &  \hspace{0.1 pt}  \hspace{0.1 pt}
                           &  \hspace{0.1 pt} \hspace{0.1 pt}
                &  \hspace{0.1 pt} \hspace{0.1 pt}
                      &  \hspace{0.1 pt}3.4\hspace{0.1 pt}                      &  \hspace{0.1 pt}\hspace{0.1 pt}
                &  \hspace{0.1 pt} 2 \hspace{0.1 pt}
                      &  \hspace{0.1 pt} $<1$ \hspace{0.1 pt}             
                \\
                $L_e-L_\mu$ no-pol~\cite{Lin:2023xyk}  \hspace{0.1 pt}  \hspace{0.1 pt}
                &  \hspace{0.1 pt}  \hspace{0.1 pt}
                &  \hspace{0.1 pt}  \hspace{0.1 pt}
                &  \hspace{0.1 pt}6\hspace{0.1 pt}
                &  \hspace{0.1 pt}  \hspace{0.1 pt}
                &  \hspace{0.1 pt} \hspace{0.1 pt}
                              &  \hspace{0.1 pt}6 \hspace{0.1 pt}
                &  \hspace{0.1 pt}  \hspace{0.1 pt}
                           &  \hspace{0.1 pt} \hspace{0.1 pt}
                &  \hspace{0.1 pt}  \hspace{0.1 pt}
                      &  \hspace{0.1 pt}\hspace{0.1 pt}                    
                      &  \hspace{0.1 pt}2.1\hspace{0.1 pt}
                          &  \hspace{0.1 pt}\hspace{0.1 pt}                        &  \hspace{0.1 pt} 2 \hspace{0.1 pt}
                      &  \hspace{0.1 pt} $<1$\hspace{0.1 pt}             
                \\    
                $L_\mu-L_\tau$ pol.~\cite{Lin:2023xyk} \hspace{0.1 pt}  \hspace{0.1 pt}
                &  \hspace{0.1 pt}  \hspace{0.1 pt}
                &  \hspace{0.1 pt} \hspace{0.1 pt}
                &  \hspace{0.1 pt}6 \hspace{0.1 pt}
                &  \hspace{0.1 pt}  \hspace{0.1 pt}
                &  \hspace{0.1 pt}  \hspace{0.1 pt}
                              &  
                \hspace{0.1 pt}6\hspace{0.1 pt}
                &  \hspace{0.1 pt}  \hspace{0.1 pt}
                           &  \hspace{0.1 pt} \hspace{0.1 pt}
                &  \hspace{0.1 pt}  \hspace{0.1 pt}
                      &  \hspace{0.1 pt}\hspace{0.1 pt}                     
                      &  \hspace{0.1 pt}3.8\hspace{0.1 pt}
                          &  \hspace{0.1 pt}\hspace{0.1 pt}                        &  \hspace{0.1 pt} 2 \hspace{0.1 pt}
                      &  \hspace{0.1 pt} $<1$\hspace{0.1 pt}             
                \\
                $L_\mu-L_\tau$ no-pol~\cite{Lin:2023xyk}  \hspace{0.1 pt}  \hspace{0.1 pt}
                &  \hspace{0.1 pt}  \hspace{0.1 pt}
                &  \hspace{0.1 pt} \hspace{0.1 pt}
                &  \hspace{0.1 pt}6  \hspace{0.1 pt}
                &  \hspace{0.1 pt}  \hspace{0.1 pt}
                &  \hspace{0.1 pt} \hspace{0.1 pt}
                              &  \hspace{0.1 pt}6 \hspace{0.1 pt}
                &  \hspace{0.1 pt}  \hspace{0.1 pt}
                           &  \hspace{0.1 pt} \hspace{0.1 pt}
                &  \hspace{0.1 pt}  \hspace{0.1 pt}
                      &  \hspace{0.1 pt}\hspace{0.1 pt}                      
                      &  \hspace{0.1 pt}0\hspace{0.1 pt}
                          &  \hspace{0.1 pt}\hspace{0.1 pt}                        &  \hspace{0.1 pt} 2 \hspace{0.1 pt}
                      &  \hspace{0.1 pt} $<1$\hspace{0.1 pt}             
                \\                                [1.0pt] \hline \hline                \end{tabular}
\caption{
We present the results of different scenarios  of the NO$\nu$A and  T2K tension. We lista when avaliable the NO and IO cases. The variable {\it Par} is the number of free parameters in each scenario. 
For the $\Delta\chi^2_{\rm \textbf{G}}$ (NO$\nu$A + T2K ) goodness-of-the-fit parameter that quantifies the quality of fit to describe experimental data and can be translated in terms of standard deviation, $\sigma$. The variable $\sigma$ quantify how much standard deviation between the two scenarios for the $\Delta\chi^2_{\rm \textbf{PG}}$ (NO$\nu$A + T2K ) parameter goodness-of-the-fit.
}
\label{tab:results}
\end{table}

\appendix

\onecolumngrid

\end{document}